\documentclass[12pt]{article}

\usepackage[dvips]{color}

\usepackage{latexsym,amsfonts}

\usepackage{cite,epsf}
\usepackage{graphicx,array}

\renewcommand{\title}[1]{\begin{center}\bf\Large #1\end{center}}

\renewcommand{\author}[1]{\begin{center}\large #1\end{center}}

\newcommand{\rr}{\mathbb{R}}
\newcommand{\zz}{\mathbb{Z}}

\def\theequation{\arabic{section}.\arabic{equation}}

\begin{document}

\begin{titlepage}
\vspace{10pt}
\hfill
{\large\bf HU-EP-09/39}
\vspace{20mm}
\begin{center}

\title{Singular Liouville fields and\\
spiky strings in $\rr^{1,2}$ and $SL(2,\rr)$}

\vspace{30pt}

{\large
George Jorjadze
}\\[5mm]

{\it
Institut f\"ur Physik der
Humboldt-Universit\"at zu Berlin,}\\
{\it Newtonstra{\ss}e 15, D-12489 Berlin, Germany}\\[2mm]
and
\\[2mm]
{\it
Razmadze Mathematical Institute,}\\
{\it M.Aleksidze 1, 0193, Tbilisi, Georgia}

\vspace{15pt}


\end{center}


\vspace{40pt}


\begin{abstract}


\noindent The closed string dynamics in $\rr^{1,2}$ and $SL(2,\rr)$
is studied within the scheme of Pohlmeyer reduction. In  both spaces two
different classes of string surfaces are specified by the structure
of the fundamental quadratic forms. The first class in $\rr^{1,2}$
is associated with the standard lightcone gauge strings and the
second class describes spiky strings and their conformal
deformations on the Virasoro coadjoint orbits. These orbits
correspond to singular Liouville fields with the monodromy matrixes
$\pm\,\, I$. The first class in $SL(2,\rr)$ is parameterized by the
Liouville fields with vanishing chiral energy functional. Similarly
to $\rr^{1,2}$, the second class in $SL(2,\rr)$ describes spiky
strings, related to the vacuum configurations of the
$SL(2,\rr)/U(1)$ coset model.

\end{abstract}

\vspace{10pt}

{\bf Keywords:} AdS-CFT correspondence, Bosonic string, Pohlmeyer reduction,\\
$~~~~~\,$Liouville equation, $SL(2,\rr)$ WZW theory, coset models.

\end{titlepage}

\noindent
{\bf \large Contents}

\vspace{2mm}

\noindent
{\bf 1$~~$ Introduction}

\vspace{2mm}

\noindent
{\bf 2$~~$ Closed strings in $\rr^{1,2}$}

\vspace{1mm}

\noindent
{2.1$~~$Pohlmeyer scheme}

\noindent
{2.2$~~$Lightcone gauge}

\vspace{1mm}

\noindent
{2.3$~~$Liouville gauge}

\vspace{2mm}

\noindent
{\bf 3$~~$  Closed strings in $SL(2,\rr)$}

\vspace{1mm}

\noindent
{3.1$~~$String dynamics in $AdS_3$}

\noindent
{3.2$~~$Map to $SL(2,\rr)$ and WZW theory}

\vspace{1mm}

\noindent {3.3$~~$Modified Pohlmeyer scheme for $SL(2,\rr)$ strings}

\vspace{1mm}

\noindent
{3.4$~~$Nilpotent gauge}

\vspace{1mm}

\noindent {3.5$~~$Liouville gauge for $SL(2,\rr)$ strings}

\vspace{2mm}

\noindent {\bf 4$~~$ Summary}

\vspace{2mm}

\noindent
{\bf 5$~~$ Appendix}


\section{Introduction}

Integrability of string dynamics in AdS spaces is one of the most
actively discussed topic of the last decade due to its important
role in the AdS/CFT correspondence. String equations have the most simple
and symmetric form in conformal coordinates.
The conformal gauge freedom of string dynamics in $AdS \times S$
can be fixed by turning the spherical part of the Virasoro constraint
to a positive constant $h$ \cite{Maldacena:2000hw}. Then, the AdS part
of the constraint becomes $-h$ and the string description splits in
two schemes of Pohlmeyer reduction \cite{pohlmeyer}. They lead to
generalized sin-Gordon and sinh-Gordon equations for the sphere and
AdS space, respectively. These equations allow a Lax pair
representation, which is a basis for the integration of string dynamics.
Details of this approach and the related list of references one can
find in \cite{gt} and \cite{Miramontes:2008wt}.

If strings propagate only in AdS space (the case $h=0$), the
described gauge fixing procedure fails. However, the Pohlmeyer
scheme can be modified in a conformally invariant form \cite{bn,
dvs, jevicki}. Recently, this approach was effectively used in
$AdS_3$, providing there a new set of interesting string solutions
\cite{Jevicki:2009uz, Alday:2009yn}.

The Pohlmeyer scheme is usually formulated in terms of a linear system
of differential equations for a basis along the string surface. This
basis is formed by the tangent and orthogonal vectors to the surface
and, therefore, the equations of the linear system contain
components of the first and the second fundamental quadratic forms.
The consistency conditions for the linear system provide dynamical
equations and chirality relations for the worldsheet variables.

In ref. \cite{Dorn:2009kq} we studied general aspects of Pohlmeyer
reduction for AdS strings in arbitrary dimensions. Motivated by the
Alday-Maldacena conjecture \cite{am}, we have extended the Pohlmeyer
scheme to the spacelike surfaces, where the chiral conditions are
replaced by holomorphic ones. To simplify the discussion, we
turned the worldsheet chiral (or holomorphic) functions to
constants. Locally this is always allowed due to the conformal
freedom. However, information about some nontrivial string
worldsheets might be encoded in global properties of chiral or
holomorphic functions (as in \cite{Alday:2009yn}) and, therefore,
such solutions could be lost by a simple gauging.

Gauge fixing in string theory is a subtle problem even in flat
spacetime. It appears that the standard light cone gauge string
surfaces in 3d Minkowski space are singular. Namely, the induced
metric tensor is degenerated at some points or lines of the surface,
and the scalar curvature diverges there.

Among the new solutions constructed in \cite{Jevicki:2009uz} there are
the spiky strings \cite{kru}, which become important ingredients
in AdS/CFT correspondence \cite{Dorey-Losi,Ishizeki:2008tx,Freyhult:2009bx}.
Note that the spiky singularities also correspond to the degeneracy
of the induced metric tensor.

A natural question related to non regular string surfaces
is to understand the character of singularities and
their role in quantized string theory.
In the present paper we study this
problem for timelike closed strings in three dimensions.

We start with the analysis of the flat case.
Using a gauge fixing for the components of the
fundamental quadratic forms, we integrate
the linear system of the Pohlmeyer scheme and realize that
the obtained surfaces are associated with the standard
lightcone gauge strings. Then we shown that the chiral $u(z)\,$
and the antichiral $\bar u(\bar z)$
components of the second quadratic form
do not have fixed signs in the lightcone gauge.

To analyze the sector with fixed signs of $u(z)\,$ and $\,\bar
u(\bar z)$ we use the gauge, which turns these functions to
constants $u(z)\mapsto \lambda,\,$ $\,\bar u(\bar z)\mapsto
\bar\lambda$. The consistency condition in this gauge reduces to the
Liouville equation, and the periodicity of closed string worldsheets
fixes the monodromy class of Liouville fields by the matrixes $\pm
I$. These are  singular Liouville fields, which are parameterized by
the Virasoro coadjoint orbits  of the vacuum configurations
$T(z)=-n^2/4$ and $\bar T(\bar z)=-\bar n^2/4$, where $n$ and $\bar
n$ are nonzero integers  \cite{BFP}. The vacuum configurations of
the Liouville field describe oscillating circular and rotating spiky
strings. The shape of a spiky string configuration at a fixed time
essentially depends on the relative sign between $\lambda$ and
$\bar\lambda$, as well as on the values of $n$ and $\bar n$. The
Virasoro coadjoint orbits define internal degrees of freedom of
spiky strings, providing  their `conformal deformations'.

In Section 3 we consider strings in $AdS_3$. Here we `improve' the
integrability of string dynamics by adding the WZ-term to the
action, which turns the system  to the  $SL(2,\rr)$ WZW theory with
Virasoro constraints \cite{Balog:1988jb}. This model, called
$SL(2,\rr)$ or $SU(1,1)$ string, was a subject of intensive study in
the 90's (see \cite{Maldacena:2000hw} for references).

We apply again the Pohlmeyer type scheme. Using the isometry between
$sl(2,\rr)$ and $\rr^{1,2}$, the scheme is formulated in a form
equivalent to 3d Minkowski case, which allows to use some results of
the previous classification here. Namely, the Kac-Moody currents
have the same parameterization as the worldsheet tangent vectors in
$\rr^{1,2}$. The next step is the integration of the Kac-Moody
currents to string worldsheets embedded in $SL(2,\rr)$. The main
difference with the flat case arises just at this point. Another
difference is related to the calculation of the worldsheet metric,
which involves the string solution, and not only the chiral
currents, as in $\rr^{1,2}$.

Finally, we summarize the results and discuss open problems.
Some technical details are shifted to the Appendixes.

\section{Closed strings in $\rr^{1,2}$}

In this section we analyze string dynamics in 3d Minkowski space
$\rr^{1,2}$ within the scheme of Pohlmeyer reduction
\cite{pohlmeyer}. This approach sheds new light to known results of
string theory in flat spacetime.

The scalar product in $\rr^{1,2}$ we denote by $X\cdot
X=X_1\,X_1+X_2\,X_2-X_0\,X_0,$ where $X_0,\,X_1,\,X_2$ are the
coordinates of $X\in \rr^{1,2}$. We consider a closed string with
periodic boundary conditions $X(\tau,\sigma+2\pi)=X(\tau,\sigma)$.
The derivatives with respect to the lightcone coordinates
$z=\tau+\sigma,\,$ $\,\bar z=\tau-\sigma$ are denoted by
$\partial\equiv\partial_z,\,$ $\,\bar\partial\equiv
\partial_{\bar z}$.

\subsection{Pohlmeyer scheme }

The  Pohlmeyer scheme for string dynamics is applied in the conformal gauge
\begin{equation}\label{conformal gauge}
\partial X\cdot\partial X=0=\bar\partial X\cdot\bar\partial X~,
\end{equation}
where $X(\tau,\sigma)\,$ satisfies the free field  equation
\begin{equation}\label{string equation}
\bar\partial\partial X=0~.
\end{equation}

The conformal gauge conditions (\ref{conformal gauge}) provide the
relation $\,\partial_\tau X\cdot\partial_\tau X=2\partial
X\cdot\bar\partial X.\,$ The timelikeness of the string worldsheet
implies $\,\partial_\tau X\cdot\partial_\tau X<0\,$ and, therefore,
the non zero component of the induced metric tensor can be
parameterized by
\begin{equation}\label{ind-metric}
\partial X\cdot\bar\partial X=-e^\alpha~.
\end{equation}
However, it has to be noted that the induced metric can be
degenerated $(\partial X\cdot\bar\partial X=0)$ at some points or
lines of the string worldsheet, where the tangent vector
$\,\partial_\tau X$ becomes lightlike. These singular points
correspond to $\alpha\rightarrow -\infty$, though the functions
$X_\mu(\tau,\sigma)$ ($\mu=0,1,2$) remain smooth (differentiable)
there.

To follow the Pohlmeyer scheme, we introduce a basis $(B,\, \bar
B,\, N)$ in $\rr^{1,2}\,$ formed by the vectors $B=\partial X,$
$\,\bar B=\bar\partial X$ and $N$, which is an unit vector
orthogonal to the string surface
\begin{equation}\label{B dot B}
N\cdot N=1~,~~~~~~B\cdot N=0=\bar B\cdot N~.
\end{equation}

Then, from (\ref{conformal gauge})-(\ref{B dot B}) one finds the
following linear system of equations for the `moving' basis along
the string world sheet
\begin{eqnarray}\label{partial B}
&&\partial B=\partial\alpha\,B+u\,N~,~~~~~~~~~~~~~~~
\bar\partial B=0~,\\  \nonumber
&&\partial \bar B=0~,
~~~~~~~~~~~~~~~~~~~~~~~~~~~~~
\bar\partial\bar B=\bar\partial\alpha\,\bar B+\bar u\,N~,\\ \nonumber
&&\partial N=e^{-\alpha}\,u\,\bar
B~,~~~~~~~~~~~~~~~~~~~~\bar\partial N=e^{-\alpha}\,\bar u\,B~,
\end{eqnarray}
where $u$ and $\bar u$ are the components of the second fundamental
form on the worldsheet
\begin{equation}\label{u}
u =\partial^2X\cdot N~,~~~~~~~~\bar u =\bar\partial^2X\cdot N~.
\end{equation}

The consistency conditions for the linear system (\ref{partial B})
are
\begin{eqnarray}\label{Gauss eq}
\bar\partial\partial \alpha+e^{-\alpha}\,u\,\bar u=0~,\\~~~~~~~
\nonumber
\partial\bar u=0~,~~~~~~
\bar\partial u=0~.
\end{eqnarray}

All these equations are invariant under the conformal transformations
\begin{eqnarray}\label{alpha mapsto}
e^{\alpha(z,\bar z)}\mapsto \zeta'(z)\bar\zeta'(\bar
z)\,e^{\alpha(\zeta(z),\bar\zeta(\bar z))}~,
~~~~~~~~~~~~~~~~~~~~~~\\ \label{u mapsto} u(z)\mapsto
\zeta'^{\,2}(z)\,u(\zeta(z))~,~~~~~~~~~~~~~\bar u(\bar z)\mapsto
\bar\zeta'^{\,2}(\bar z)\,\bar u(\bar\zeta(\bar z))~,~~~~~~
\end{eqnarray}
together with $X(z,\bar z)\mapsto X\left(\zeta(z),\bar\zeta(\bar
z)\right)$ and $N(z,\bar z)\mapsto N\left(\zeta(z),\bar\zeta(\bar
z)\right)$. The functions $\zeta(z)$ and $\bar\zeta(\bar z)$ here
are monotonic and they satisfy the monodromy conditions
\begin{equation}\label{monodromy of zeta}
\zeta(z+2\pi)=\zeta(z)+2\pi~,~~~~~~~~\bar\zeta(\bar
z+2\pi)=\bar\zeta(\bar z)+2\pi~.
\end{equation}
This symmetry is a remnant of the reparameterization invariance of
string theory in the conformal coordinates. One can use this
invariance to remove remaining non physical degrees of freedom and
simplify the integration procedure.

We follow this scheme in the next two subsections. The functions
$u(z)$ and $\bar u(\bar z)$ are assumed smooth and periodic, like
the components of the tangent vectors $B_\mu(z)$ and $\bar
B_\mu(\bar z)$. We specify two different classes of $u(z),\,$ $\bar
u(\bar z)$. The first class is formed by the functions which change
signs in the interval of periodicity, whereas $u(z)$ and $\bar
u(\bar z)$ have no zeros for the second class. The gauge
fixing conditions differ for these classes are different.
After integration of the
linear system (\ref{partial B}) we realize that the first class
corresponds to the standard lightcone gauge and the second class
describes spiky and oscillating circular strings.

\subsection{Lightcone gauge}

The scheme proposed in this subsection is quite similar to the one
used in \cite{Barbashov} and \cite{Jevicki:2009uz}. Before
integration of the linear system (\ref{partial B}) we have to find
solutions of the consistency conditions (\ref{Gauss eq}). These
conditions are satisfied by the following simple parameterization
\begin{eqnarray}\label{u=f'}
u(z)=f'(z)~,~~~~~~\bar u(\bar z)=-\bar f'(\bar z)~,~~~~~~~~
e^\alpha=\frac{1}{2}\,\big[\bar f(\bar z)-f(z) \big]^2~.
\end{eqnarray}
The aim is to describe the class of functions $f$ and $\bar f,$
which lead to periodic $X(\tau,\sigma)$.
Note that solutions of the linear system (\ref{partial B}) with
periodic coefficients, in general, are only quasi-periodic.
Therefore, the periodicity of the functions $u,$ $\bar u$ and
$e^\alpha$ is a necessary, but not a sufficient condition for
periodicity of $X(\tau,\sigma)$.

The integration of the linear system (\ref{partial B}) with  $u,\,$
$\bar u$ and $\alpha$ given by (\ref{u=f'}) is done in Appendix A
and it leads to
\begin{equation}\label{B=f}
B=f(z)\,{\bf e}+{\bf e}_{_{+}}+f^2(z)\,{\bf e}_{_{-}}~,~~~~~~ \bar
B=\bar f(\bar z)\,{\bf e}+{\bf e}_{_{+}}+\bar f^2(\bar z)\,{\bf
\nonumber e}_{_{-}}~,
\end{equation}
\begin{equation}\label{N=f}
N=\frac{\bar f(\bar z)+f(z)}{\bar f(\bar z)-f(z)}\,\,{\bf e}
+\frac{2}{\bar f(\bar z)-f(z)}\,\,{\bf e}_{_{+}}+\frac{2\, f(z)
\,\bar f(\bar z)}{\bar f(\bar z)-f(z)}\,\,{\bf e}_{_{-}} ~.
\end{equation}
Here ${\bf e},$ $\,{\bf e}_{_{+}},$ $\,{\bf e}_{_{-}}$ are $(z, \bar
z)$-independent $\rr^{1,2}$-valued vectors, which arise as
integration constants in solving (\ref{partial B}). The map from the
vectors $({\bf e},{\bf e}_{_{+}},{\bf e}_{_{-}})$ to $(B,\bar B, N)$
is invertible and the orthonormality conditions for the basis
$(B,\bar B, N)$ is equivalent to
\begin{equation}\label{e.e}
{\bf e}\cdot{\bf e}=1~,~~~~~{\bf e}\cdot{\bf e}_{_{+}}=0={\bf
e}\cdot{\bf e}_{_{-}}={\bf e}_{_{+}}\cdot{\bf e}_{_{+}}={\bf
e}_{_{-}}\cdot{\bf e}_{_{-}}~,~~~~~~{\bf e}_{_{+}}\cdot{\bf
e}_{_{-}}=-\,\frac{1}{2}~.
\end{equation}
These conditions are realized by
\begin{equation}\label{e=e_a}
{\bf e}_{_{\pm}}=\frac{1}{2}\,({\bf e}_{_{0}}\pm{\bf e}_{_{1}})~,
~~~~~~~~{\bf e}={\bf e}_{_{2}}~,
\end{equation}
where ${\bf e}_\mu$ $(\mu=0,1,2)$ is the standard orthonormal basis
in $\rr^{1,2}$, with ${\bf e}_\mu^{~\nu}=\delta_\mu^{~\nu}.$ The
scalar and exterior products of these basis vectors
\begin{equation}\label{e_a.e_b}
{\bf e}_\mu\cdot {\bf e}_\nu
=\eta_{\mu\nu}=\mbox{diag}(-1,1,1)~,~~~~~~{\bf e}_\mu\times{\bf
e}_\nu=\epsilon_{\mu\nu}\,^\rho\,{\bf e}_\rho~,
\end{equation}
are given by the metric $\eta_{\mu\nu}$  and the Levi-Civita
$\epsilon_{\mu\nu\rho}$ tensors, respectively. With
$\epsilon_{012}=1$, the algebra of exterior products yields ${\bf
e}\times{\bf e}_{_{\pm}}=\pm\,{\bf e}_{_{\pm}}$ and $2{\bf
e}_{_{-}}\times{\bf e}_{_{+}}={\bf e}.$ Using then (\ref{u=f'}), the
normal vector (\ref{N=F bar F}) can be written in a Lorentz
invariant form
\begin{equation}\label{N=B tims B}
N=e^{-\alpha}\,\bar B\times B~. \nonumber
\end{equation}
Other realizations of (\ref{e.e}) are obtained by Lorentz
transformations of (\ref{e=e_a}).

The tangent vectors to a closed string worldsheet $B=\partial X$ and
$\bar B=\bar\partial X$ are periodic chiral and antichiral vector
functions respectively. Hence, the functions $f(z)$ and $\bar f(\bar
z)$ in (\ref{B=f}) have to be periodic  as well and they  enjoy the
Fourier mode expansions
\begin{equation}\label{Fourier modes}
f(z)=p+\sum_{n\neq 0}a_n\,e^{-inz}~,~~~~~~~~~\bar f(\bar z)=\bar
p+\sum_{n\neq 0}\bar a_n\,e^{-in\bar z}~.
\end{equation}
In addition, the periodicity of $X(\tau,\sigma)$ requires equality
of the zero modes of $B(z)\,$ and $\,\bar B(\bar z).$ These
conditions are given by
\begin{equation}\label{p=bar p}
p=\bar p~,~~~~~~~~~~~\sum_{n>0} |a_n|^2=\sum_{n>0} |\bar a_n|^2~.
\end{equation}
Eqs. (\ref{Fourier modes})-(\ref{p=bar p}) define the class of
$f(z),$ $\,\bar f(\bar z)$ leading to periodic $X(\tau,\sigma)$.
These string surfaces, besides $f(z)\,$ and $\,\bar f(\bar z),\,$
depend on the parameters of Lorentz transformations of the basis
(\ref{e=e_a}) and also on three integration constants related to the
final equations $\partial X=B$ and $\bar\partial X=\bar B$.

The general solution of the consistency conditions (\ref{Gauss eq}),
given by \cite{Jevicki:2009uz}
\begin{equation}\label{g-solution}
e^\alpha=\frac{\big[\Phi(z) +\bar\Phi(\bar
z)\big]^2}{2\Phi'(z)\,\bar\Phi\,'(\bar z)}\, \,u(z)\,\bar u(\bar
z)~,
\end{equation}
depends on two chiral ($u,\,\Phi$) and two antichiral ($\bar
u,\,\bar\Phi$) functions. The parameterization (\ref{u=f'}) defines
a `constraint surface' in the space of fundamental quadratic forms
and it can be treated as a gauge fixing condition. In fact, a gauge
fixing condition in (\ref{g-solution}) can be written in the form
\begin{equation}\label{gauge fixing}
\frac{u(z)}{\Phi'(z)}=a=\frac{\bar u(\bar z)}{\bar\Phi'(\bar z)}~,
\end{equation}
where $a$ is a constant. Then, with $f(z)=a\Phi(z)$ and $\bar f(\bar
z)=-a \bar\Phi(\bar z)$ we obtain (\ref{u=f'}).

In order to find independent parameterizing variables of string
surfaces $X(\tau,\sigma),\,$ it is important to analyze the
remaining freedom of conformal transformations in (\ref{u=f'}). This
analysis is done in Appendix B. It shows that the freedom of
conformal transformations in (\ref{u=f'}) is described by three
parameters $\phi_0,$ $\bar\phi_0$ and $c$. The first two correspond
to translations in the chiral and antichiral sectors. The
transformations parameterized by $c$ are $f,$  $\bar f$ dependent.
Their infinitesimal form is defined by
\begin{equation}\label{small zeta'}
\zeta'(z)=1+\varepsilon\,c (f(z)-p)~,~~~~~~~~~\bar\zeta'(\bar
z)=1+\varepsilon\,c(\bar f(\bar z)-p)~.
\end{equation}
The variable $c$ could be included in the infinitesimal parameter
$\varepsilon$, however, it is more convenient to keep this form. In
Appendix B we also show that the conditions (\ref{p=bar p}) are
invariant under these conformal transformations. We use the
remaining conformal symmetry to reduce the number of parameterizing
variables.

Let's consider Lorentz transformations of the basis (\ref{e=e_a}). A
boost in $X_1$-direction transforms the basis $({\bf
e}_{_{+}},\,{\bf e}_{_{-}},\,{\bf e})$ to $(P\,{\bf
e}_{_{+}},\,P^{-1}\,{\bf e}_{_{-}},\,{\bf e})$, where $P>0$ and
$\theta=\log P$ is the boost parameter. The corresponding tangent
vectors (\ref{B=f}) become $P$ dependent
\begin{equation}\label{B=f,P}
B=f\,{\bf e}+P\,{\bf e}_{_{+}}+f^2\,P^{-1}\,{\bf e}_{_{-}}~,~~~~~~
\bar B=\bar f\,{\bf e}+P\,{\bf e}_{_{+}}+\bar f^2\,P^{-1}\,{\bf
\nonumber e}_{_{-}}~.
\end{equation}
Below we show that this equation defines the general form of the
tangent vectors. Namely, further Lorentz transformations of the
basis (\ref{e=e_a}) correspond either to transformed parameterizing
variables ($f(z),\,\bar f(\bar z), P$), or to the remaining conformal
freedom.

We divided infinitesimal Lorentz transformations in three
independent groups
\begin{eqnarray}\label{e mapsto 1}
&&1.~~~{\bf e}_{_{+}}\mapsto {\bf e}_{_{+}}+\varepsilon {\bf
e}_{_{+}}~,~~~ {\bf e}_{_{-}}\mapsto {\bf e}_{_{-}}- \varepsilon
{\bf e}_{_{-}}~, ~~~{\bf e}\mapsto {\bf e}~;\\ \label{e mapsto 2}
&&2.~~~{\bf e}_{_{+}}\mapsto {\bf e}_{_{+}}+\varepsilon {\bf
e}~,~~~~~ {\bf e}_{_{-}}\mapsto {\bf e}_{_{-}}~,~~~~~~~~~~~{\bf
e}\mapsto {\bf e}+2\varepsilon {\bf e}_{_{-}}~; \\\label{e mapsto 3}
&&3.~~~{\bf e}_{_{+}}\mapsto {\bf e}_{_{+}}~,~~~~~~~~~~~~ {\bf
e}_{_{-}}\mapsto {\bf e}_{_{-}}+\varepsilon{\bf e} ~~~~~~~{\bf
e}\mapsto {\bf e}+2\varepsilon{\bf e}_{_{+}}~.
\end{eqnarray}
Eq. (\ref{e mapsto 1}) corresponds to a boost in $X_1$-direction,
whereas (\ref{e mapsto 2}) and (\ref{e mapsto 3}) are two different
compositions of a boost in  $X_2$-direction and a rotation in
$(X_1,X_2)$-plane.

The transformations (\ref{e mapsto 1}) and (\ref{e mapsto 2})
preserve the structure of the tangent vectors (\ref{B=f,P}) with
transformed ($f,\,\bar f,\,P$). From (\ref{e mapsto 1}) and
(\ref{B=f,P}) one gets
\begin{equation}\label{f mapsto 1}
f\mapsto f~,~~~~~~~~~~~~\bar f\mapsto\bar f~,~~~~~~P\mapsto
P+\varepsilon P~,
\end{equation}
and similarly (\ref{e mapsto 2}) leads to
\begin{equation}\label{f mapsto 2}
f\mapsto f+\varepsilon P~,~~~~~~\bar f\mapsto\bar f+\varepsilon
P~,~~~~~P\mapsto P~.
\end{equation}
Eq. (\ref{f mapsto 1}) is consistent with the definition of $P$, as a
boost parameter in $X_1$-direction, and eq. (\ref{f mapsto 2})
states that the Lorentz transformations (\ref{e mapsto 2})
correspond to translations of the zero modes of $f(z)$ and $\bar
f(\bar z)$. This means that the parameters related to the
transformations (\ref{e mapsto 1}) and (\ref{e mapsto 2}) can be
neglected, since the corresponding freedom is encoded in dilatations
of $P$ and translations of $f$ and $\bar f$.

The transformations (\ref{e mapsto 3}) are of different type. They
change the tangent vector $B$ in (\ref{B=f,P}) in the following way
\begin{eqnarray}\label{B mapsto}
B\mapsto B_\varepsilon=\left(f+\varepsilon\,f^2\,P^{-1}\right)\,{\bf
e}+\left(P+2\varepsilon\,f\right)\,{\bf e}_{_{+}}+f^2\,P^{-1}\,{\bf
e}_{_{-}}~.
\end{eqnarray}
This destroys the structure of (\ref{B=f,P}). In particular, the
${\bf e}_{_{+}}$- components of the transformed tangent vector
$B_\varepsilon$ is not constant anymore. The same is valid for $\bar
B_\varepsilon$.

Here we use the remaining conformal freedom (\ref{small zeta'}). An
infinitesimal transformation $z\mapsto z+\varepsilon\phi(z)$
corresponds to
\begin{equation}\label{B-epsilon} B_\varepsilon(z)\mapsto
(1+\varepsilon\phi'(z))\left[B_\varepsilon(z)+\varepsilon\,\phi(z)
B_\varepsilon'(z)\right]~,
\end{equation}
and the coefficient of ${\bf e}_{_{+}}$ becomes constant with
\begin{equation}\label{phi'=}
\phi'(z)=-2\,P^{-1}\,(f(z)-p)~,
\end{equation}
which is an allowed conformal transformation (\ref{small zeta'})
with $c=-2\,P^{-1}$. The transformed constant coefficient of ${\bf
e}_{_{+}}$ is equal to $P+2\varepsilon p$ and it is easy to check
that the transformed coefficients of ${\bf e}_{_{-}}$ and ${\bf e}$
are related as in (\ref{B=f,P}).

Summarizing the discussion on Lorentz transformations of the basis
(\ref{e=e_a}), we conclude that eq. (\ref{B=f,P}) indeed describes
the general form of the tangent vectors. 

From (\ref{B=f,P}) follows
that $\partial_\tau X_+=P$ and $\partial_\sigma X_+=0,\,$ where
$X_+$ is the ${\bf e}_{_{+}}$ component of $X(\tau,\sigma)$.  These
are the standard light cone gauge conditions in 3d bosonic
string. 

The above mentioned integration constants of the equations
$\partial X=B$ and $\bar\partial X=\bar B,$ together with the
freedom in ($\phi_0,$ $\bar\phi_0$)-translations, describe the
coordinate zero modes of $X(\tau,\sigma)$ in the lightcone gauge.
Thus, the parameterization of the first and the second fundamental
forms by (\ref{u=f'}), after factorization of the remaining
conformal symmetry, corresponds to the lightcone gauge.

Now we analyze the conformal factor of the metric tensor $e^\alpha$,
defined by (\ref{u=f'}). The function $\bar f-f$ is given as a sum
of the non-zero Fourier modes
\begin{equation}\label{f+bar f}
\bar f-f=\sum_{n\neq 0}\Big[\bar a_n\,e^{-in(\tau-\sigma)}
-a_n\,e^{-in(\tau+\sigma)}\Big]~,
\end{equation}
and its integration by $\sigma$ around the unit circle vanishes.
This means that $\bar f(\bar z)-f(z)$ has not a fixed sign. The
points where this function vanishes correspond to the above
mentioned degeneracy of the induced metric, i.e. $\partial_\tau
X\cdot\partial_\tau X=0=\partial_\sigma X\cdot\partial_\sigma X$ and
$\alpha\rightarrow -\infty$. Calculating the tangent vector
$\partial_\sigma X=B-\bar B$, from (\ref{B=f}) we find
\begin{equation}\label{X'}
\partial_\sigma X=(f-\bar f)\,\,[\,{\bf e}+(f+\bar f)\,{\bf
e}_{_{-}}\,]~,
\end{equation}
which vanishes at $\bar f(\bar z)-f(z)=0$. Note that the normal
vector (\ref{N=f}) diverges at these points . Since this vector has
the unit norm, it diverges in the lightlike direction.

The worldsheet scalar curvature, calculated in the conformal
coordinates, is given by
$R=-2e^{-\alpha}\,\bar\partial\partial\,\alpha$. In the
parameterization (\ref{u=f'}), it takes the form
\begin{equation}\label{R}
R=-\frac{8\,f'(z)\,\bar f'(\bar z)}{[\bar f(\bar z)-f(z)]^4}~,
\end{equation}
which is singular at $\bar f(\bar z)-f(z)=0$. This singularity can not
be removed by coordinate transformations. Hence, 
the lightcone gauge string surfaces in
three dimensions are always singular.

In higher dimensions, the conformal factor of the induced metric
tensor in the lightcone gauge is given by
\begin{equation}\label{e^alpha for d>3}
e^\alpha=\frac{1}{2}\sum_a[\bar f_a(\bar z)-f_a(z)]^2~,
\end{equation}
where the summation index $a$ corresponds to the transverse (to
$\,{\bf e}_{_{+}}\,$ and $\,{\bf e}_{_{-}}$) coordinates. Each $\bar
f_a-f_a$ has the structure (\ref{f+bar f}) and, therefore, they
vanish at some points. But if these points for different $a$'s do
not coincide, $\alpha$ is globally regular.

\vspace{3mm}

The Fourier modes in (\ref{Fourier modes}) are canonical variables,
which are used for the quantization of the lightcone bosonic string
\cite{GSW}. The key point for a consistent quantization is to check
the commutation relations of the Poincare group generators. This
calculation in an arbitrary dimension of spacetime is non-trivial
only for the commutators of the Lorentz transformations
$\,[M_{-\,a},M_{-\,b}],\,$ where $\,a\,$ and $\,b\,$ are indices for
the transverse coordinates. The Poincare symmetry requires vanishing
of these commutator, which in 3 dimensions is trivially fulfilled,
since there is only one transverse coordinate. So, there is no
quantum anomaly in the Poincare algebra of the lightcone quantized
3d bosonic string. However, it appears that there is an additional
class of string solutions in three dimensions, which is not covered
by the lightcone gauge strings.

\vspace{3mm}

Before introducing the new class, we describe those properties of
$u$ and $\bar u$, which distinguish the
classes.
These functions have vanishing zero modes
\begin{equation}\label{zero modes=0}
\int_0^{2\pi}\mbox{d}z\,\,u(z)=0=\int_0^{2\pi}\mbox{d}\bar z\,\,\bar
u(\bar z)~,
\end{equation}
since $u=f'$, $\bar u=\bar f'$ and  $f,$ $\bar f$ are periodic. Note
that if $u=0$, $\,\bar u=0$, the tangent vectors (\ref{B=f,P})
become constants and the string surface degenerates to a massless
particle trajectory. Neglecting this degenerated case, from
(\ref{zero modes=0}) follows that $u(z)$ and $\bar u(\bar z)$
change signs in the interval of periodicity. This property is
obviously invariant under the conformal transformations (\ref{alpha
mapsto}).

Thus, the lightcone gauge describes the string surfaces with
changing signs of $u(z)$ and $\bar u(\bar z)$. In the next
subsection we show that the class of string surfaces with fixed signs of
$u(z)$ and $\bar u(\bar z)$ is not empty, and this class describes 
oscillating circular and rotating spiky strings.

There is an additional class of $u(z),$ $\bar u(\bar z)$, 
which have zeros, but do not change signs there. 
It is an `intermidiate' class between
the lightcone and spiky strings.
The corresponding surfaces have different type of singularities, which
`move' in the lightcone directions around the $(\tau,\sigma)$-cylinder.
We do not consider this class in this paper.

\subsection{Liouville gauge}

Suppose $u(z)$ and $\bar u(\bar z)$ have no zeros. Such functions
can be transformed to constants $u(z)\mapsto\lambda,\,$ $\,\bar
u(\bar z)\mapsto\bar\lambda$ by the conformal transformation (\ref{u
mapsto}). The dynamical variables $\lambda$ and $\bar\lambda$ have
the same signs as $u$ and $\bar u$, respectively, and their modules
are given by the conformal invariants
\begin{equation}\label{lambda}
2\pi\sqrt{|\lambda|}=\int_0^{2\pi}dz\,\sqrt{|u(z)|}~,~~~~~~~
2\pi\sqrt{|\bar\lambda|}=\int_0^{2\pi}d\bar z\,\sqrt{|\bar u(\bar
z)|}~,
\end{equation}
which easily follow from the monodromy properties of $\zeta$ and
$\bar\zeta$.

The choice of constant $u(z)\,$ and $\,\bar u(\bar z)$ fixes the
conformal gauge freedom up to zero modes of $\zeta$ and $\bar\zeta$.
We call this choice the Liouville gauge, since the corresponding
consistency condition (\ref{Gauss eq}) reduces to the Liouville
equation
\begin{equation}\label{Liouville eq}
\partial\bar\partial\alpha+\lambda\bar\lambda\,e^{-\alpha}=0~.
\end{equation}
The general solution of this equation is given by
\begin{equation}\label{Liouv-g-solution}
e^{-\alpha}=\frac{2}{|\lambda\bar\lambda|}\,\,\frac{F'(z)\,\bar
F'(\bar z)}{\big[\epsilon F(z)+\bar\epsilon\bar F(\bar z)\big]^2} ~,
\end{equation}
where $F,$ $\,\bar F$ are monotonic functions  $F'>0$, $\,\bar F'>0$
and $\epsilon=\mbox{sign}\,\,\lambda,$
$\bar\epsilon=\mbox{sign}\,\,\bar\lambda.$ For a symmetry reason, we
treat all four possibilities of ($\epsilon$, $\bar\epsilon$)
simultaneously, though the general solution (\ref{Liouv-g-solution})
depends only on the sign of $\epsilon\bar\epsilon$.

The integration of the linear system (\ref{partial B})
with $\alpha$ given by (\ref{Liouv-g-solution}) and
$u(z)=\lambda,$ $\bar u(\bar z)=\bar\lambda$ can be done similarly
to the lightcone gauge. Repeating the same steps as before (see
Appendix A) we obtain
\begin{eqnarray}\nonumber
&&B=\frac{\lambda}{F'(z)}\,\big[\,\,\,F(z)\,\,\,{\bf
e}\,+\,\epsilon\,{\bf e}_{_{+}}\,+\,\epsilon\,F^2(z)\,{\bf
e}_{_{-}}\big] ~,\\ \label{B=F}
&&\bar B =
\frac{\bar\lambda}{\bar F'(\bar z)}\big[-\bar F(\bar z)\,{\bf
e}\,+\,\bar\epsilon\,{\bf e}_{_{+}}\,+\,\bar\epsilon\,\bar F^2(\bar
z)\,{\bf e}_{_{-}}\big]~,\\ \label{N=F bar F}
&&N=\frac{\bar\epsilon\,\bar F(\bar z)-\epsilon\,
F(z)}{\epsilon\,F(z)+\bar\epsilon\,\bar F(\bar z)}\,{\bf e}\,-
\,\frac{2}{\epsilon\,F(z)+\bar\epsilon\,\bar F(\bar z)}\,\,{\bf
e}_{_{+}}\,+\,\frac{2\,\bar\epsilon\,\epsilon\,\,\bar F(\bar
z)\,F(z)}{\epsilon\,F(z)+\bar\epsilon\,\bar F(\bar z)}\,\,{\bf
e}_{_{-}}~,
\end{eqnarray}
Here $({\bf e},{\bf e}_{_{+}},{\bf e}_{_{-}})$ are again
$\rr^{1,2}$-valued integration constants with the same
orthonormality conditions (\ref{e.e}).

The space of Liouville fields (\ref{Liouv-g-solution}) is invariant
under the transformations
\begin{equation}\label{Liouville conformal map}
e^{-\alpha(z,\bar z)}\mapsto \zeta'(z)\bar\zeta'(\bar
z)\,\,e^{-\alpha(\zeta(z),\bar\zeta(\bar z))}~,
\end{equation}
which  corresponds to $F(z)\mapsto F(\zeta(z)),\,$ $\bar F(\bar
z)\mapsto \bar F(\bar\zeta(\bar z))$. These are the conformal
transformations in Liouville theory. In spite of similarity, there
is an essential differences between the conformal transformations
(\ref{alpha mapsto}) and (\ref{Liouville conformal map}). Namely,
the transformations (\ref{alpha mapsto}) describe the freedom in
choice of conformal coordinates and they do not change the string
surface. Whereas, (\ref{Liouville conformal map}) acts on the
Liouville fields and it changes the date of the linear system
(\ref{partial B}), which is not a worldsheet reparameterization.
Note also that the conformal weight of $e^{-\alpha}$ is equal to $1$
by (\ref{Liouville conformal map}) and $-1$ by (\ref{alpha mapsto}).
To avoid misunderstanding with these two conformal transformations,
we use for (\ref{Liouville conformal map}) and the related maps in
Liouville theory the name Virasoro transformations.

Let us discuss the regularity issue of closed string worldsheets in
the Liouville gauge, related to peculiarities of the Liouville field
$\alpha(\tau,\sigma)$. It is well known that a globally regular
Liouville field on a cylindrical spacetime exist only for
$\epsilon\bar\epsilon=-1$ and it belongs to the hyperbolic monodromy
\cite{PP}. The parameterizing functions of this monodromy class
satisfy the conditions $F(z+2\pi)=e^P\,F(z)$ and $\bar F(\bar
z+2\pi)=e^P\,\bar F(\bar z)$, with $P>0$.

However, these conditions do not correspond to periodic tangent
vectors (\ref{B=F}). It means that the linear system (\ref{partial
B}) does not provide a closed string configuration in the Liouville
gauge, if the Liouville field on the cylinder is regular.

There is a class of singular Liouville fields with a regular stress
tensor of the theory and some other remarkable properties
\cite{JPP,Marn,BFP}. Singularities of these fields correspond to
zeros of the exponent $e^{\alpha}$. The equation
$e^{\alpha(\tau,\sigma)}=0$ is equivalent to $\epsilon
F(z)+\bar\epsilon\bar F(\bar z)=0$ and it defines non-intersecting,
smooth lines on the ($\tau,\, \sigma$)-manifold. It appears that
this type of singular Liouville fields on the cylinder can match the
periodicity conditions of closed string dynamics.

The authors of ref. \cite{BFP} gave a complete classification of
periodic Liouville fields by the coadjoint orbits of the Virasoro
algebra. This classification is based on the analysis of the
Schr\"odinger (Hill) equation  with a periodic potential given by
the stress tensor of Liouville theory. We use these results here
to select an appropriate class of Liouville fields. In Appendix C we
give a list of relations in Liouville theory which are helpful for
understanding of the technical details below.

The solutions of Hill equations (\ref{Hill eq}) in the chiral and
the antichiral sectors are denoted by $\psi(z),\,\chi(z)$ and
$\bar\psi(\bar z),\,\bar\chi(\bar z)$, respectively. They are
normalized by the unit Wronskians (\ref{Wronskians}). The
parameterization of the functions $F(z)$ and $\,\bar F(\bar z)$ in
terms of these solutions (\ref{F=}) define the following form of the
tangent vectors (\ref{B=F})
\begin{eqnarray}\label{B=psi}
&&B=\lambda\epsilon[~\psi(z)\,\,\chi(z)\,\,{\bf
e}+\psi^2(z)\,{\bf e}_{_{+}}+\chi^2(z)\,{\bf e}_{_{-}}]~,\\
\label{bar B=bar psi} \nonumber &&\bar B
=\bar\lambda\bar\epsilon[-\bar\psi(\bar z)\,\bar\chi(\bar z)\,{\bf
e}+\bar\chi^2(\bar z)\,{\bf e}_{_{+}}+\bar\psi^2(\bar z)\,{\bf
e}_{_{-}}]~.
\end{eqnarray}
The periodicity of conditions $B(z+2\pi)=B(z),$ $\,\bar B(\bar
z+2\pi)=\bar B(\bar z)$ requires
\begin{eqnarray}\label{monodromy E}
&&\psi(z+2\pi)=\pm\psi(z)~,~~~~~~~\chi(z+2\pi)=\pm\chi(z)~,\\
\nonumber &&\bar\psi(\bar z+2\pi)=\pm\bar\psi(\bar
z)~,~~~~~~~\bar\chi(\bar z+2\pi)=\pm\bar\chi(\bar z)~,
\end{eqnarray}
which corresponds to the monodromy matrix $\pm\,I$.
In the classification
of the coadjoint orbits this class is denoted by $E_\pm$.
Its typical representatives are
\begin{eqnarray}\label{ground state}
\psi_k(z)=\sqrt{\frac{2}{k}}\, \cos\left(\frac{kz}{2}\right)~,~~~~~~
\chi_k(z)=~\epsilon\,\sqrt{\frac{2}{k}}\,
\sin\left(\frac{kz}{2}\right)~, \\ \nonumber
\bar\psi_{\bar k}(\bar z)=\sqrt{\frac{2}{\bar
k}}\,\cos\left(\frac{\bar k\bar z}{2}\right)~, ~~~~~~ \bar\chi_{\bar
k}(\bar z)=-\bar\epsilon\,\sqrt{\frac{2}{\bar
k}}\,\sin\left(\frac{\bar k\bar z}{2}\right)~,
\end{eqnarray}
where $k$ and $\bar k$ are positive integers. They count the number
of zeros of these functions in the interval $[0,2\pi)$. The
coefficients in front of $\sin$ and $\cos$-functions correspond to
the normalization of Wronskians (\ref{Wronskians}). The
corresponding Liouville field configurations are associated with
vacuum solutions, since the stress tensor for these fields is
constant
\begin{eqnarray}\label{vacuum T}
T(z)=-\frac{k^2}{4}~,~~~~~\bar T(\bar z)=-\frac{\bar k^2}{4}~.
\end{eqnarray}
The general representatives of $E_\pm,\,$ are obtained by the
Virasoro transformations of the functions (\ref{ground state}) with
the conformal weight $-\frac{1}{2}$
\begin{eqnarray}\label{psi mapsto psi}
&&\psi_k(z)\mapsto
\left(\zeta'(z)\right)^{-\frac{1}{2}}\,\psi_k(\zeta(z))~, ~~~~~~~~~~
\,\chi_k(z)\mapsto
\left(\zeta'(z)\right)^{-\frac{1}{2}}\,\chi_k(\zeta(z))~,\\ \nonumber
&&\bar\psi_k(\bar z)\mapsto
\left(\bar\zeta'(\bar z)\right)^{-\frac{1}{2}}\,
\bar\psi_k(\bar\zeta(\bar z))~, ~~~~~~~~~~
\bar\chi_k(\bar z)\mapsto
\left(\bar\zeta'(\bar z)\right)^{-\frac{1}{2}}\,
\bar\chi_k(\bar\zeta(\bar z))~.
\end{eqnarray}
Thus, the acceptable class of Liouville fields is associated with
the Virasoro group orbits of the vacuum configurations.

Let us consider the string configurations related to (\ref{ground
state}) in more detail. In this case the tangent vectors
(\ref{B=psi}) read
\begin{eqnarray}\label{B=sin-cos}
&&B(z)=\Lambda\,[(1+\cos(nz)) \,{\bf e}_{_{+}}+(1-\cos(nz))\, {\bf
e}_{_{-}}+\sin(nz)\,\,{\bf e}]~,
\\
 \nonumber
&&\bar B(\bar z)=\bar\Lambda\,[(1-\cos(\bar n\bar z))\,{\bf
e}_{_{+}}+ (1+\cos(\bar n\bar z))\,{\bf e}_{_{-}}+\sin(\bar n\bar
z)\,{\bf e}]~,
\end{eqnarray}
where we have introduced the notations
\begin{equation}\label{notations 1}
\Lambda=\frac{|\lambda|}{k}~,~~~~~\bar\Lambda=\frac{|\bar\lambda|}{\bar
k}~,~~~~~n= \epsilon k~,~~~~\bar n= \bar\epsilon\bar k~.
\end{equation}
Note that $\Lambda$ and $\bar\Lambda$ are positive and $n$, $\bar n$
are nonzero integers. It is easy to see that the periodicity of
$X(\tau,\sigma)$ requires $\Lambda=\bar\Lambda$.

To proceed, we choose the constant basis vectors as in (\ref{e=e_a})
\begin{equation}\label{basis}
{\bf e}=\left(\begin{array}{cr} 0\\0\\1
\end{array}\right)~,~~~~~~
{\bf e}_{_{+}}=\frac{1}{2}\left(\begin{array}{cr}
1\\1\\0\end{array}\right)~,~~~~~~{\bf e}_{_{-}}=
\frac{1}{2}\left(\begin{array}{cr}
~\,\,1\\-1\\\,\,~0\end{array}\right)~,
\end{equation}
and rewrite eq. (\ref{B=sin-cos}) in the form
\begin{equation}\label{B=()}
B=\Lambda
\left(\begin{array}{cr} 1\\ \cos(nz)\\
\sin(nz)\end{array} \right)~,~~~~~~ \bar B
=\Lambda\left(\begin{array}{cr} 1\\-\cos(\bar n\bar z)\\
~~~\sin (\bar n\bar z)\end{array}\right)~.
\end{equation}
The integration of these tangent vectors, up to translations, yields
the string surface
\begin{equation}\label{X=()}
X(\tau,\sigma)=\frac{\Lambda}{n\bar n}\, \left(\begin{array}{cr}
2n\bar n\tau\\~~\bar n\sin(n\tau+n\sigma)~-
~n\sin(\bar n\tau-\bar n\sigma)\\
-\bar n\cos(n\tau+n\sigma)~-~n\cos(\bar n\tau- \bar
n\sigma)\end{array} \right)~.
\end{equation}
Thus, the time component is proportional to $\tau$
\begin{equation}\label{X-0=}
X^0=2\Lambda\tau~,
\end{equation}
and the spatial part is described by the function
\begin{equation}\label{Z(tau,sigma) in R}
Z(\tau,\sigma)=\frac{\Lambda}{in}\,\,e^{i(n\tau+n\sigma)}+
\frac{\Lambda}{i\bar n}\,\,e^{-i(\bar n\tau-\bar n\sigma)}~,
\end{equation}
where $Z=X_1+iX_2$ is the complex coordinate on the
$(X_1,X_2)$-plane.

The conformal factor (\ref{ind-metric}) of the induced metric is
given by
\begin{equation}\label{e^alpha=}
e^{\alpha}=\Lambda^2\,[1+\cos(nz+\bar n\bar z)]~,
\end{equation}
and this function vanishes at
\begin{equation}\label{zeros of e^alpha}
(n+\bar n)\tau+(n-\bar n)\sigma=(2m+1)\pi~, ~~~~~~~~~~~ (m\in \zz)~.
\end{equation}
The zeros of $e^\alpha$ correspond to $\partial_\tau
X\cdot\partial_\tau X=0=\partial_\sigma X\cdot\partial_\sigma X.$
From eq. (\ref{B=()}) follows that the vector  $\partial_\tau
X=B+\bar B$ is indeed lightlike at the singular points (\ref{zeros
of e^alpha}) and  $\partial_\sigma X=B-\bar B$  vanishes there. The
conditions $\partial_\sigma X=0$ and the lightlikeness of
$\partial_\tau X$ are the boundary conditions for an open string.
Therefore, the singular points on the worldsheet look like the end
points of an open string  and they have a spiky character.

According to (\ref{zeros of e^alpha}), the number of spikes for a
fixed $\tau$ is equal to $|n-\bar n|$. The case $\bar n=n$ is
special and we consider it separately.

The string worldsheet (\ref{X=()}) for $\bar n=n$ reduces to
\begin{equation}\label{X=()+}
X(\tau,\sigma)=\frac{2\Lambda}{n}\, \left(\begin{array}{cr}
n\tau\\~~~\cos(n\tau)\,\,
\sin(n\sigma)\\
-\cos(n\tau)\,\,\cos(n\sigma)\end{array}
\right)~.
\end{equation}
The corresponding string configuration at a fixed $\tau$ is a circle
on the ($X_1,\,X_2$)-plane with the center at the origin. The radius
of the circle oscillates in time. At $\,\tau=
(m+\frac{1}{2})\frac{\pi}{n}\,,\,$ $\,m\in Z,\,$ the circle shrinks
to the origin. Thus, the case $\,\bar n=n\,$ describes oscillating
circular strings without spikes.

Now we consider the general case with an arbitrary $n$ and $\bar n$,
except $\bar n=n$.

From eq. (\ref{Z(tau,sigma) in R}) follows the relation
\begin{equation}\label{X1-X_2=1}
Z(\tau,\sigma)=e^{i\omega\tau}\,Z(\sigma+\omega_0\tau)~,
\end{equation}
where
\begin{equation}\label{spiky frequency}
\omega=-\frac{2n\bar n}{n-\bar n}~,~~~~~~~~ \omega_0=\frac{n+\bar
n}{n-\bar n}~,
\end{equation}
and $Z(\sigma)=Z(0,\sigma)$ corresponds to the string configuration
at $\tau=0$. The shift of the argument $\sigma$ by $\omega_0\tau$
does not change the string shape. Therefore, eq. (\ref{X1-X_2=1})
describes a rotating string around the origin with the frequency
$\omega$. The sign of $\omega$ defines the direction of the
rotation. Thus, the initial shape is preserved in dynamics.

Using again (\ref{X=()}), the shape of the string configuration at
$\tau=0$ can be written as
\begin{equation}\label{X1-X_2=2}
Z(\sigma)=\frac{\Lambda}{in}\,\,e^{in\sigma}+\frac{\Lambda}{i\bar
n}\,\,e^{i\bar n\sigma}~.
\end{equation}
This curve, given as a composition of two `rotations', is known as a
epicycloid if $n$ and $\bar n$ have the same sign and a hypocycloid
if their signs are opposite. The rotation parameter is $\sigma$. The
integers $n$ and $\bar n$ define the rotation frequencies and the
rotation radiuses are given by their inverse numbers, up to  the
scale factor $\Lambda$.

In Appendix D we give several plots of $Z(\sigma)$ for different
values of $n$ and $\bar n$. Below we list some properties of the
curves (\ref{X1-X_2=2}), which help to understand the structure of
these plots and also to visualize the general case.

{\bf 1.} The spikes at $\tau=0$ correspond to the points
\begin{equation}\label{sigma_m}
\sigma_m=\frac{2m+1}{|n-\bar n|}\,\pi~,~~~~~~~~~
m=0,\,1,...,\,|n-\bar n|-1~,
\end{equation}
and they are located on the circle with the center at the origin and
the radius
\begin{equation}\label{spiky distance}
|Z(\sigma_m)|=\frac{\Lambda\,|n-\bar n|}{|n\bar n|}~.
\end{equation}

{\bf 2.} The following inequalities hold for $\sigma\in
(\sigma_m,\sigma_{m+1})$
\begin{eqnarray}\label{string distance}
\frac{|n+\bar n|}{|n\bar n|}\leq\frac{|Z(\sigma_m)|}{\Lambda}<
\frac{|n-\bar n|}{|n\bar n|}~,~~~~~~~\mbox{for}~~~ n\bar n<0~;\\
\nonumber \frac{|n-\bar n|}{|n\bar
n|}<\frac{|Z(\sigma_m)|}{\Lambda}\leq \frac{|n+\bar n|}{|n\bar
n|}~,~~~~~~~\mbox{for}~~~ n\bar n>0~.
\end{eqnarray}

{\bf 3.} The spike corresponding to $\sigma_m$ has the direction of
the radius vector  at $Z(\sigma_m)\,$ if $\,n\bar n<0$; and the
direction of the spike is opposite to the radius vector, if $\,n\bar
n>0$.

{\bf 4.} The curvature of $Z(\sigma)$, $\sigma\in
(\sigma_m,\sigma_{m+1})$, with respect to the origin is positive, if
$\,n\bar n>0$; it is zero, if $n+\bar n=0$; and it is negative, if
$n\bar n<0$ and $n+\bar n\neq 0$.

{\bf 5.} When $\sigma$ changes from $\sigma_m$ to $\sigma_{m+1}$,
the polar angle of $Z(\sigma)$ rotates on
\begin{equation}\label{delta phi}
\Delta\phi=2\pi\,\frac{\mbox{min}(|n|,|\bar n|)}{|n-\bar n|}~.
\end{equation}

According to (\ref{spiky distance}) and (\ref{string distance}), the
spikes are the farthest points from the origin, if $n\bar n<0$, and
they are the nearest ones, if $n\bar n>0$. These two properties
easily follow from eq. (\ref{X1-X_2=2}). The proof of other
properties and some additional information about the spiky strings
is given in Appendix E. The derivation of eq. (\ref{delta phi})
there does not apply to the case $n+\bar n=0$ (see eq. (\ref{delta
phi=})). We consider this case here separately.

If $n+\bar n=0$, eq. (\ref{X=()}) provides the surface
\begin{equation}\label{X=()-}
X(\tau,\sigma)=\frac{2\Lambda}{n}\, \left(\begin{array}{cr} n\tau\\
\cos(n\tau)\,
\sin(n\sigma)\\
\sin(n\tau)\,\sin(n\sigma)\end{array} \right)~,
\end{equation}
which describes a folded ($n$-times) rotating string. The spikes are
at $\sigma_m=(m+\frac{1}{2})\frac{\pi}{|n|},\,$ $m=0,1,...,2|n|-1.$
They correspond to one end of the folded string for even $m$, and to
another end for odd $m$. Thus, one gets $\Delta\phi=\pi$, which is
consistent with (\ref{delta phi}).

Eq. (\ref{delta phi}) indicates that the curve $Z(\sigma)$ is
non-intersecting if $\mbox{min}(|n|,|\bar n|)=1$ and $|n|\neq |\bar
n|$. For example, if $\bar n=-1$ and $n>1$, one gets the spiky
strings of \cite{kru}
\begin{equation}\label{X=()-1}
X(\tau,\sigma)=\frac{\Lambda}{n}\, \left(\begin{array}{cr}
2n\tau\\~~\sin(n\tau+n\sigma)
-n\sin(\tau-\sigma)\\
-\cos(n\tau+n\sigma)+n\cos(\tau-\sigma)\end{array} \right)~.
\end{equation}
In general, hypocycloids and epicycloids are intersecting (or
folded) curves.

The properties 1 - 5 help to visualize these curves and draw them
qualitatively. The properties 1 and 3 provide the positions of
spikes and their directions, respectively. Other properties (2, 4
and 5) define how the spikes are connected by smooth curves. The
form of these curves and the direction of the spikes essentially
depend on the sign of $n\bar n$, as one can see in Appendix D.

Since these curves correspond to a composition of two rotations,
they arise in description of simple mechanical systems. Their
properties were investigated long time ago and one can find them in
the literature. Here we present them for completeness.

The string solutions given as rotating hypocycloids and epicycloids
first were obtained in \cite{Burden:1982zb} as a model of hadrons.
Later these solutions were rediscovered in cosmic strings and in
AdS/CFT correspondence by different authors. A list of references
and interesting comments one can find in a recent letter paper
\cite{Burden:2008zz}.

\vspace{3mm}

The general solution in the Liouville gauge is obtained by the
Poincare and Virasoro transformations of the vacuum solutions
(\ref{X=()}). The Lorentz transformation in the Poincare group
correspond to the freedom in choice of the basis $({\bf e},{\bf
e}_{_{+}},{\bf e}_{_{-}})$ and the translations are related to the
integration constants, which we have neglected in (\ref{X=()}). The
Virasoro transformations (\ref{psi mapsto psi}) map the tangent
vectors (\ref{B=()}) to
\begin{equation}\label{B=()1}
B(z)=\frac{\Lambda}{\zeta'(z)}\,
\left(\begin{array}{cr} 1\\ \cos(n\zeta(z))\\
\sin(n\zeta(z))\end{array} \right)~,~~~~~~\bar B (\bar z)
=\frac{\Lambda}{\bar\zeta'(\bar z)}\left(\begin{array}{cr} 1\\
-\cos(\bar n\bar\zeta(\bar z))\\
~~\sin (\bar n\bar\zeta(\bar z))\end{array}\right)~.
\end{equation}
To integrate these vectors to a periodic $X(\tau,\sigma)$, their
zero modes have to be equal
\begin{equation}\label{B=bar B}
\int_0^{2\pi}dz\,B(z)= \int_0^{2\pi}d\bar z\,\bar B(\bar z)~.
\end{equation}
This equation relates $\zeta\,$ and $\,\bar\zeta$ by three (one for
each vector component) conditions.

The induced metric now is degenerated at $n\zeta(z)+\bar
n\bar\zeta(\bar z)=(2m+1)\pi\,\,$ ($m\in Z$) and the tangent vector
$\zeta'(z)\partial X-\bar\zeta'(\bar z)\bar\partial X$ vanishes
there.

Let us consider the following infinitesimal transformation
\begin{equation}\label{small zeta 1}
\zeta(z)=z+\varepsilon_0+\frac{\varepsilon_1}{n}\,\sin(nz)+
\frac{\varepsilon_2}{n}\,\cos(nz)~.
\end{equation}
It is easy to check that the corresponding $B(z)$ in (\ref{B=()1})
is a Lorentz transformed vacuum vector from (\ref{B=()}). In
particular, $\varepsilon_0$ becomes an infinitesimal rotation angle
in ($X_1, X_2$) plane, whereas $\varepsilon_1$ and $\varepsilon_2$
are infinitesimal boost parameters in $X_1$ and $X_2$ directions,
respectively. Thus, the Virasoro transformations of the
parameterizing Liouville field contain the Lorentz transformations
of the target space as a subgroup.

As it was mentioned in the previous subsection, the lightcone gauge
does not cover the string solutions of the Liouville gauge.
Therefore, the complete quantum picture of $\rr^{1,2}$ strings
requires quantization of singular Liouville fields, which is an open
and a challenging problem in its own right.

\setcounter{equation}{0}

\section{Closed strings in $SL(2,\rr)$}

We start this section with a standard approach to string dynamics in
AdS spaces. Then we pass from $AdS_3$ to $SL(2,\rr)$ group valued
variables and introduce the chiral structure of WZW theory there.
This enables us to formulate the Pohlmeyer type scheme in the same
manner as for $\rr^{1,2}$.

\subsection{String dynamics in $AdS_{3}$}

 $AdS_{3}$ is realized as the hyperbola
\begin{equation}\label{hyperbola}
Y\cdot Y+1=0
\end {equation}
embedded in $\rr^{2,2}$. $\,Y\equiv (Y^{\tilde0},Y^0;Y^1,X^2)$
denotes a point in $\rr^{2,2}$ and the scalar product $Y\cdot Y=Y^M
Y^N\,G_{MN}$ is defined by the metric tensor
$G_{MN}=\mbox{diag}(-1,-1;1,1)$.

The choice of conformal coordinates on a timelike  string worldsheet
$Y(\tau,\sigma)$ assumes
\begin{equation}\label{conformal gauge1}
\partial Y\cdot\partial Y=0=\bar\partial Y\cdot\bar\partial Y~,
\end{equation}
and the nonzero element of the induced metric tensor is
parameterized as in (\ref{ind-metric})
\begin{equation}\label{ind-metric1}
\partial Y\cdot\bar\partial Y=-e^\alpha~.
\end{equation}
String dynamics in the conformal gauge is described by the
Lagrangian
\begin{equation}\label{Lagrangian}
{\mathcal L}=\partial Y\cdot \bar\partial Y +\Lambda(Y\cdot Y+1)~,
\end{equation}
where $\Lambda$ is a Lagrange multiplier. Its elimination from the
equation of motion yields
\begin{equation}\label{string equation1}
\bar\partial\partial Y+e^\alpha\,Y=0~.
\end{equation}
The Pohlmeyer scheme for this system leads to the $\sinh$-Gordon
equation for the $\,\alpha\,$ field \cite{bn, dvs}. Though the
system is formally integrable,  one can write in an explicit form
only the string solutions corresponding to the $\sinh$-Gordon
solitons \cite{Jevicki:2009uz} (see also \cite{Sakai:2009ut}).
To improve the
integrability of the system by the structure of WZW theory
\cite{W1}, we use the isometry between $AdS_3$ and the $SL(2,\rr)$
group manifold.

\subsection{Map to $SL(2,\rr)$ and WZW theory}

First we describe the isometry between the $sl(2,\rr)$ algebra and
$\rr^{1,2}$. Let us introduce the basis in $sl(2,\rr)$
\begin{equation}\label{sl2 basis}
  t_0=\left( \begin{array}{cr}
  0&1\\-1&0 \end{array}\right)~,~~~~
   t_1=\left( \begin{array}{cr}
  0&~1\\1&~0 \end{array}\right)~,~~~~
 t_2=\left( \begin{array}{cr}
  1&0\\0&-1 \end{array}\right)~.
\end{equation}
These three matrices ($t_\mu,$ $\mu=0,1,2)$ satisfy the relations
\begin{equation}\label{tt=}
t_\mu\,t_\nu=\eta_{\mu\nu}\, I+\epsilon_{\mu\nu}\,^\rho\,t_\rho,
\end{equation}
where $\eta_{\mu\nu}=\mbox{diag}(-1,1,1)$ form the metric tensor of
3d Minkowski space, $I$ denotes the unit matrix and
$\epsilon_{\mu\nu\rho}$ is the Levi-Civita tensor with
$\epsilon_{012}=1$ (see (\ref{e_a.e_b})). The expansion of $a\in
sl(2,\rr)$ in the basis (\ref{sl2 basis}), $\,a=a^\mu\,t_\mu,\,$
provides a map $a\mapsto a^\mu$ from $sl(2,\rr)$ to $\rr^{1,2}$. The
inner product in $sl(2,\rr)$, introduced by the normalized trace
$\langle a\,b\rangle=\frac{1}{2}\,\mbox{tr}(a\,b),\,$ leads to
$\,\langle t_\mu\,t_\nu\rangle=\eta_{\mu\nu}$ and makes this map
isometric.

A helpful remark is in order here. The transformations of the
adjoint representation $a\mapsto g\,a\,g^{-1}\,$ ($g\in SL(2,\rr)$)
preserve the inner product in $sl(2,\rr)$. Therefore, the matrixes
$\Lambda^\mu\,_\nu=\langle \,t^\mu\,g\,t_\nu\,g^{-1}\,\rangle$
define the Lorentz transformations of $\rr^{1,2}$. Since $SL(2,\rr)$
is connected, $\langle \,t^\mu\,g\,t_\nu\,g^{-1}\,\rangle\in
SO_{\uparrow}(1,2)$ and $\langle \,t^0\,g\,t_0\,g^{-1}\,\rangle\geq
1.$

Now we consider the map from $Y\in AdS_3$ to $g\in SL(2,\rr)$
\begin{equation}\label{g=Y}
  g=\left( \begin{array}{cr}
 Y^{\tilde0}+Y^2 &Y^1+Y^0\\Y^1-Y^0& Y^{\tilde0}-Y^2
 \end{array}\right)~,
\end{equation}
which provides the equivalence between eq. (\ref{hyperbola}) and the
condition $\mbox{det}\,g=1$. Eq. (\ref{g=Y}) can be written in the
form $g=Y^{\tilde0}\,I+Y^\mu\,t_\mu.$ Therefore, the inverse map is
given by
\begin{equation}\label{Y=g}
Y^{\tilde0}=\langle\,g\,\rangle~,~~~~~~~Y^\mu=\langle\,t^\mu\,
g\,\rangle~.
\end{equation}
Note that $g^{-1}=Y^{\tilde0}\,I-Y^\mu\,t_\mu$. Using these compact
forms of $g$ and $g^{-1}$ in terms of $t_\mu$ matrices, and the
algebraic relations (\ref{tt=}), one easily checks that
\begin{equation}\label{dg=dY}
\langle\, g^{-1}\,dg\,\,g^{-1}\,d g\rangle= dY\cdot dY~.
\end{equation}
Due to this isometry, the conformal gauge conditions (\ref{conformal
gauge1}) are equivalent to
\begin{equation}\label{conformal gauge2}
\langle\, g^{-1}\,\partial g\,\,g^{-1}\,\partial g\rangle=0= \langle
g^{-1}\,\bar\partial g\,\,g^{-1}\,\bar\partial g\rangle~,
\end{equation}
and the parameterization of the nonzero component of the worldsheet
metric by (\ref{ind-metric1}) can be written as
\begin{equation}\label{ind-metric2}
\langle\, g^{-1}\,\partial g\,g^{-1}\,\bar\partial g\rangle
=-e^\alpha~.
\end{equation}
The $AdS_3$ string dynamics in the conformal gauge is described by
the action
\begin{equation}\label{action-0}
S_0=\int dz\,d\bar z\,\langle\, g^{-1}\,\partial
g\,\,g^{-1}\,\bar\partial g\rangle~.
\end{equation}
Its variation leads to the equation of motion
\begin{equation}\label{g-eq}
\partial\left(g^{-1}\,\bar\partial g\right)+
\bar\partial\left(g^{-1}\,\partial g\right)=0~,
\end{equation}
which corresponds to (\ref{string equation1}), together with
(\ref{hyperbola}) and (\ref{ind-metric1}).

To get the equations of WZW theory \cite{W1}
\begin{equation}\label{WZW-eq}
\bar\partial(\partial g\,g^{-1}) = 0=\partial(g^{-1}\,\bar \partial
g)~,
\end{equation}
one has to add to the action (\ref{action-0}) the WZ-term, which is
a volume integral of the 3-form $H=\frac{1}{3}\,\langle
g^{-1}\mbox{d}g \wedge g^{-1}\mbox{d}g \wedge g^{-1}\mbox{d}g
\rangle$. This form on $SL(2,\rr)$ is exact $H=\mbox{d}F$, with
\begin{equation}\label{2-form}
 F =\frac{\langle t_0~ g^{-1}~\mbox{d}g\rangle \wedge
\langle t_0~ \mbox{d}g~g^{-1}\rangle} {1+ \langle t^{\,0}~
g~t_0~g^{-1} \rangle}~.
\end{equation}
Note that this 2-form is globally well defined due to the remark
above. Then, with Stokes' theorem, the action of the $SL(2,\rr)$ WZW
theory is given by a surface integral from the Lagrangian \cite{FJW}
\begin{equation}\label{WZW-Lagrangian}
\mathcal L_{WZW}=\langle g^{-1}\partial g~ g^{-1}\bar\partial
g\rangle+ \frac{\langle t_0~ g^{-1}\partial g~\rangle\langle t_0~
\bar\partial g~g^{-1} \rangle -\langle t_0~ g^{-1}\bar\partial
g\rangle \langle t_0 ~\partial g~g^{-1} \rangle} {1+\langle
t^{\,0}~g~t_0~g^{-1}\rangle}~.
\end{equation}
The Euler-Lagrange equations obtained from (\ref{WZW-Lagrangian})
reproduce the equations of WZW theory (\ref{WZW-eq}). Adding to
these equations the conformal gauge conditions (\ref{conformal
gauge2}), one gets a system called the $SL(2,\rr)$ string
\cite{Balog:1988jb}.

In the next subsection we investigate this system
by the Pohlmeyer scheme.

\subsection{Pohlmeyer scheme for $SL(2,\rr)$ string}

Before starting the Pohlmeyer scheme note that a reparameterization
invariant description of the system, yielding both the equation of
motion (\ref{WZW-eq}) and the constraints (\ref{WZW-eq}), is given
by the action
\begin{eqnarray}\label{action}
S=-\,\frac{1}{2}\,
\int d^2\xi\,\Big[\sqrt{|h|}\,\,h^{ab}\langle g^{-1}\partial_a
g~ g^{-1}\,\partial_b g\rangle +
~~~~~~~~~~~~~~~~~~~~~~~~~~~~~~~~~~~~~~~~~~~
\\ \nonumber \epsilon^{ab}\,\,\frac{\langle t_0~ g^{-1}\partial_a
g~\rangle\langle t_0~
\partial_b g~g^{-1} \rangle -\langle t_0~ g^{-1}\partial_b
g\rangle \langle t_0 ~\partial_a g~g^{-1} \rangle} {1+\langle
t^{\,0}~g~t_0~g^{-1}\rangle}\Big]~.~~~
\end{eqnarray}
Here $\xi^a\,$ $\,(a=0,1)$ are worldsheet coordinates, $h$ is the
determinant of the worldsheet metric tensor $h_{ab}$, $h^{ab}$ is
its inverse and $\epsilon^{ab}$ is the 2d Levi-Civita tensor with
$\epsilon^{01}=1$. In the conformal gauge
$h_{ab}\sim\mbox{diag}\,(-1,1)$, we indeed obtain eqs.
(\ref{conformal gauge2}) and (\ref{WZW-eq}), with $\xi^0=\tau$ and
$\xi^1=\sigma$.

Let us consider the Kac-Moody currents
\begin{equation}\label{KM-currents}
J=\partial g\,g^{-1}~,~~~~~~~~~\bar J=g^{-1}\,\bar\partial g~,
\end{equation}
which, according to (\ref{WZW-eq}), satisfy the chirality conditions
\begin{equation}\label{KM chirality}
\bar\partial J=0=\partial\bar J~.
\end{equation}
The parameterization of the induced metric (\ref{ind-metric2}) in
terms of these currents reads
\begin{equation}\label{ind-metric3}
\langle\, J\, g\,\bar J\,g^{-1}\,\rangle=-e^\alpha~,
\end{equation}
and the conformal gauge conditions (\ref{conformal gauge2}) are
\begin{equation}\label{J^2=o}
\langle \,J\,J\rangle=0=\langle \,\bar J\,\bar J\rangle~.
\end{equation}

Due to the isometry between $sl(2,\rr)$ and $\rr^{1,2}$, $J(z)$ and
$\bar J(\bar z)$ are associated with lightlike vectors as
$B(z)=\partial X$ and
$\bar B(\bar z)=\bar\partial X$ in $\rr^{1,2}$.

Similarly to the $\rr^{1,2}$ case, we consider the inner product of
the Kac-Moody currents $\langle\, J\, \bar J\,\rangle$. Taking into
account that $J$ and $\bar J$ are lightlike and $g\,\bar J\,g^{-1}$
corresponds to a proper Lorentz transformation of $\bar J$, one
finds that $\langle\, J\, \bar J\,\rangle$ and $\langle\, J\,
g\,\bar J\,g^{-1}\,\rangle$ have the same sign. Therefore, we can
use the parameterization
\begin{equation}\label{e^beta}
\langle\, J\, \bar J\,\rangle=-e^\beta~.
\end{equation}
Thus, the data for the Kac-Moody currents $(J,\,\bar J)$ and the
tangent vectors $(B,\,\bar B)$ are  similar. Using then the isometry
between $sl(2,\rr)$ and $\rr^{1,2}$, one gets the same linear system
for a moving basis as (\ref{partial B})
\begin{eqnarray}\label{partial J}
&&\partial J=\partial\beta\,J+v\,K~,~~~~~~~~~~~~~~~~
\bar\partial J=0~,\\ \nonumber
&&\partial \bar J=0~,
~~~~~~~~~~~~~~~~~~~~~~~~~~~~~
\bar\partial\bar J=\bar\partial\beta\,\bar J+\bar v\,K~,\\ \nonumber
&&\partial K=e^{-\beta}\,v\,\bar
J~,~~~~~~~~~~~~~~~~~~~~\bar\partial K=e^{-\beta}\,\bar v\,J~.
\end{eqnarray}
Here $K$ is a $sl(2,\rr)$ valued unit vector, orthogonal to $J$ and
$\bar J$
\begin{equation}\label{KK,KJ}
\langle\, K\, K\,\rangle=1~,~~~~~~\langle\, K\, J\,\rangle=0=
\langle\, K\, \bar J\,\rangle~,
\end{equation}
and $v,$ $\bar v$ are defined similarly to the coefficients of the
second fundamental form (\ref{u})
\begin{equation}\label{v}
v =\langle\,\partial J\, K\,\rangle~,
~~~~~~~~\bar v =\langle\,\bar\partial J\, K\,\rangle~.
\end{equation}
The consistency conditions for the linear system (\ref{partial J}),
\begin{eqnarray}\label{Gauss eq beta}
\bar\partial\partial \beta + e^{-\beta}\,v\,\bar v=0~,
~~~~~~~
\partial\bar v=0~,~~~~~~
\bar\partial v=0~,
\end{eqnarray}
coincide with (\ref{Gauss eq}).

Due to the equivalence with the $\rr^{1,2}$ case, we use the same
gauge fixing conditions. The first corresponds to the
parameterization
\begin{eqnarray}\label{v=h'}
v(z)=f'(z)~,~~~~~~\bar v(\bar z)=-\bar f'(\bar z)~,~~~~~~~~
e^\beta=\frac{1}{2}\,\big[\bar f(\bar z)-f(z)\big]^2~,
\end{eqnarray}
and the second  to the Liouville gauge with
\begin{equation}\label{Liouville gauge}
v(z)=\lambda~,~~~~~\bar v(\bar z)=\bar \lambda~,~~~~~
\bar\partial\partial\beta+\lambda\lambda\,e^{-\beta}~.
\end{equation}
As we will see in the next subsection the conditions (\ref{v=h'})
describe the nilpotently gauged WZW theory.

On the level of solutions of the linear system, the pair $(J, \bar
J)$ is  completely equivalent to $(B, \bar B)$, and we can use the
solutions (\ref{B=f}) and (\ref{B=psi}).

The next step is a construction of string worldsheets $g(z,\bar z)$.
The WZW field  splits into the product of chiral (left) and
anti-chiral (right) fields
\begin{equation}\label{WZ-field}
g(z,\bar z)=g_l(z)\,g_r(\bar z)~,
\end{equation}
and to find the string worldsheet one has to integrate the equations
\begin{equation}\label{g'=jg}
g_l'(z)=J(z)\,g_l(z)~,~~~~~~~~~~g_r'(\bar z)=g_r(\bar z)\,\bar
J(\bar z)~.
\end{equation}
Then the induced metric (\ref{ind-metric3}) can be calculated by
\begin{equation}\label{ind-metric4}
e^\alpha=-
\langle\, g_l^{-1}(z)\, g_l'(z)\,g'_r(\bar z),g_r^{-1}(\bar z)\,\rangle~.
\end{equation}
We realize this programme in the following two subsections. Some
helpful formulas for $SL(2,\rr)$ calculations are presented in
Appendix F.

\subsection{Nilpotent gauge }

Let us consider the gauge (\ref{v=h'}). The isometry between
$sl(2,\rr)$ and $\rr^{1,2}$ relates the standard orthonormal bases
of this spaces $t_\mu\leftrightarrow {\bf e}_\mu,$ $\,\mu=(0,1,2)$.
The basis (\ref{e=e_a}) then corresponds to
\begin{equation}\label{e mapsto t}
{\bf e} \leftrightarrow t_2=\left( \begin{array}{cr} 1&0\\
0&-1\end{array}\right)~,~~~~~~~
{\bf e}_{_{+}} \leftrightarrow t_+=\left( \begin{array}{cr} 0&1\\
0&0\end{array}\right)~,~~~~~~~
{\bf e}_{_{-}} \leftrightarrow t_-=\left( \begin{array}{cr} ~0&0\\
-1&0\end{array}\right)~,
\end{equation}
and similarly to (\ref{B=f}) we obtain the Kac-Moody currents
\begin{eqnarray}\label{J=fe+}
J(z)=\left( \begin{array}{cr} ~\,\,f(z)&~1~\,\\
\\
-f^2(z)&-f(z)\end{array}\right)~, ~~~~~~~~~~ \bar J(\bar z)=\left(
\begin{array}{cr} ~\,\,\bar f(\bar z)&~1~\,
\\ \\
-\bar f^2(\bar z)&-\bar f(\bar z)\end{array}\right)~.
\end{eqnarray}

The matrixes $t_\pm=\frac{1}{2}(t_0\pm t_1)$ are nilpotent elements
$(t_\pm^2=0)$ of the $sl(2,\rr)$ algebra. The currents (\ref{J=fe+})
have constant components in $t_+$ direction equal to $1$
($J^{+}=1=\bar J^{+})$. Note that the transformation to the basis
$(P\,{\bf e}_{_{+}},\,P^{-1}\,{\bf e}_{_{-}},\,{\bf e})$ used in
(\ref{B=f,P}) is equivalent to the rescaling of the $t_+$ and $t_-$
components in (\ref{J=fe+}) by $P$ and $P^{-1}$, respectively.

It is well known that the nilpotent gauging of the $SL(2,\rr)$ WZW
model leads to Liouville theory \cite{Dublin}. This gauging
corresponds to the Hamiltonian reduction with constant $J^{+}(z)$
and $\,\bar J^{+}(\bar z),$ similarly to (\ref{J=fe+}). The
Kac-Moody currents (\ref{J=fe+}) satisfy the Virasoro constraints
(\ref{J^2=o}) as well. Constant $J^{+}(z),$ $\,\bar J^{+}(\bar z)$
and the Virasoro constraints together form the second class
constraints. Thus, these two sets of constraints are complementary
to each other and they provide a coset construction.

Applying the reduction scheme used in coset WZW models, we write the
chiral and antichiral parts of the WZW-field in a matrix form
\begin{eqnarray}\label{g_l,g_r}
g_l(z)=\left( \begin{array}{cr} \,\psi(z)& \,\chi(z)~\\ \\
\xi(z)&\eta(z)\end{array}\right)~, ~~~~~~~~~~
g_r(\bar z)=\left( \begin{array}{cr} ~\bar\xi(\bar z)&\bar\psi(\bar z)
\\  \\
\bar\eta(\bar z)&\bar\chi(\bar z)\end{array}\right)~,
\end{eqnarray}
and from (\ref{g'=jg})  find the relations
\begin{eqnarray}\label{psi_1}
&&\xi(z)=\psi'(z)-f(z)\,\psi(z)~,~~~~~~~~
\bar\xi(\bar z)=\bar\psi'(\bar z)+\bar f(\bar z)\,\bar\psi(\bar z)~,\\
\nonumber &&\eta(z)=\chi'(z)-f(z)\,\chi(z)~,~~~~~~~~ \bar\eta(\bar
z)=\chi'(\bar z)+\bar f(\bar z)\,\bar\chi(\bar z)~,
\end{eqnarray}
and the Hill equations
\begin{eqnarray}\label{Hill(2)}
&&\psi''(z)=f'(z)\,\psi(z)~,~~~~~~~~
\bar\psi''(\bar z)=-\bar f'(\bar z)\,\bar\psi(\bar z)~,\\
\nonumber &&\chi''(z)=f'(z)\,\chi(z)~,~~~~~~~~\, \bar\chi''(\bar
z)=-\bar f'(\bar z)\,\bar\chi(\bar z)~,
\end{eqnarray}
which are satisfied  by the components of the first row of
$~~g_l(z)~$ and the second column of $~g_r(\bar z)$. These
components are invariant under the gauge transformations generated
by the nilpotent currents. The unimodularity conditions
$\mbox{det}\,g_l(z)=1=\mbox{det}\,g_r(\bar z)$ provide the unit
Wronskians, as the normalization conditions for the solutions of
(\ref{Hill(2)})
\begin{equation}\label{Wronskians(2)}
\psi(z)\chi'(z)-\psi'(z)\,\chi(z)=1~,~~~~~~ \bar\psi(\bar
z)\bar\chi'(\bar z)-\bar\psi'(\bar z)\, \bar\chi(\bar z)=-1~.
\end{equation}

Thus, we get the WZW-field
\begin{eqnarray}\label{nilpotent solutions}
g=\left( \begin{array}{cr} \,\psi(z)& \,\chi(z)~\\ \\
\psi'(z)-f(z)\psi(z)&\chi'(z)-f(z)\chi(z)\end{array}\right) \left(
\begin{array}{cr} ~\bar\psi'(\bar z)+\bar f(\bar z)
\bar \psi(\bar z)&\bar\psi(\bar z)
\\  \\
\bar\chi'(\bar z)+\bar f(\bar z)\bar\chi(\bar z)& \bar\chi(\bar
z)\end{array}\right),
\end{eqnarray}
parameterized by the gauge invariant chiral ($\psi,\,\chi$) and
antichiral ($\bar\psi,\,\bar\chi$) fields, which are related by the
unit Wronskians (\ref{Wronskians(2)}). The $g_{12}$ matrix element
of the WZW-field $g_{12}(z,\bar z)=\psi(z)\bar\psi(\bar
z)+\chi(z)\bar\chi(\bar z)$ is also gauge invariant and it is
identified with the Liouville field exponent $V(z,\bar z)$ of the
conformal weight $-\frac{1}{2}$ (see Appendix C). The potentials in
the Hill equations form the stress tensor of Liouville theory
\begin{equation}\label{f'=T}
f'(z)=T(z)~,~~~~~~~~~-\bar f'(\bar z)=\bar T(\bar z)~.
\end{equation}

On the other hand, the WZW-field (\ref{nilpotent solutions})
describes a string surface in $SL(2,\rr)$. The worldsheet induced
metric (\ref{ind-metric4}) obtained from (\ref{nilpotent solutions})
reads
\begin{equation}\label{induced metric(5)}
e^\alpha=\left[\psi'(z)\,\bar\psi'(\bar z)+\chi'(z)\,\bar\chi'(\bar
z)\right]^2~.
\end{equation}

In this way we parameterize the $SL(2,\rr)$ string surfaces by the
chiral and antichiral functions of Liouville theory. However, the
fact that the components of the stress tensor (\ref{f'=T}) are given
as derivatives of periodic functions imposes certain restrictions on
allowed Liouville fields. Namely, the chiral and antichiral energy
functionals of Liouville theory, given by the integral of the stress
tensor over the period, have to vanish
\begin{equation}\label{E=0}
L_0=\int_0^{2\pi}\mbox{d}z\,\, T(z)=0~,~~~~~~\bar
L_0=\int_0^{2\pi}\mbox{d}\bar z\,\, \bar T(\bar z)=0~.
\end{equation}

Note that solutions of the Hill equation with a periodic potential
are only quasi-periodic. Writing the pairs ($\psi,\,\chi$) and
$\bar\psi,\,\bar\chi$) as a row and column, respectively
\begin{equation}\label{Monodromy of psi}
\Psi^T=(\psi~~\chi)~,~~~~~~~~
\bar\Psi=\left( \begin{array}{cr} \bar\psi \\
\bar\chi\end{array}\right)~,
\end{equation}
one gets $\Psi^T(z+2\pi)=\Psi^T(z) M$ and  $\bar\Psi(\bar
z+2\pi)=\bar M\bar\Psi(\bar z),$ with $M\in SL(2,\rr)$ and $\bar
M\in SL(2,\rr).$ The monodromies of the chiral WZW-fields in
(\ref{nilpotent solutions}) then are given by
\begin{equation}\label{Monodromy(2)}
g_l(z+2\pi)=g_l(z)\,M~,~~~~~~~g_r(\bar z+2\pi)=\bar M\,g_r(\bar z)
\end{equation}
and the periodicity of $g(\tau,\sigma)$ requires $M=\bar M$. The
transformation $g_l(z)\mapsto g_l(z)N$, $g_r(\bar z)\mapsto
N^{-1}g_r(\bar z)$ leaves the general solution (\ref{WZ-field})
invariant and transforms the monodromy matrix by $M\mapsto
N^{-1}MN$. The monodromies with $|\langle M\rangle|< 1,\,$
$\,|\langle M\rangle|=1\,$ and $\,|\langle M\rangle|>1$ are called
elliptic, parabolic and hyperbolic, respectively. The monodromy
properties of the chiral fields ($\psi, \chi$) play an important
role in the classification of Liouville fields in terms of the
coadjoint orbits of the Virasoro algebra \cite{BFP}.

As an illustrative toy example of the scheme let's consider the case
with constant $f(z)$ and $\bar f(\bar z)$. They correspond to
vanishing $T(z)$ and $\bar T(\bar z),$ and the solutions of the Hill
equations with the unit Wronskians are
\begin{equation}\label{parabolic}
\psi (z)=1~,~~~~~\chi(z)=z~;~~~~~~~\bar\psi(\bar z)=\bar
z~,~~~~~\bar\chi(\bar z)=1~.
\end{equation}
These functions correspond to the parabolic monodromy with
$M=I+2\pi\,t_+$. Inserting them in (\ref{nilpotent solutions}) with
$f(z)=c$ and $\bar f(\bar z)=\bar c$ one gets a WZW-field, which
depends only on $\tau$, and describes a particle trajectory like
constant $f(z)$ and $\bar f(\bar z)$ in $\rr^{1,2}$.

In general, the Hill equation can not be integrated explicitly for
an arbitrary potential. Therefore, in contrast to $\rr^{1,2}$, the
functions $f$ and $\bar f$ are not convenient parameterizing
variables for string surfaces in $SL(2,\rr).$  Usually, it is more
helpful to parameterize the functions $(\psi,\,\chi)$ and
($\bar\psi, \bar\chi$) directly and express $f$ and $\bar f$ through
them.

Let's consider the following class of chiral and antichiral fields
\begin{eqnarray}\label{elliptic solutions}
\psi(z)=\frac{\cos[\theta\zeta(z)]}{\sqrt{\theta\,\zeta'(z)}}~,~~~~~~~~~
\bar\psi(\bar z)=\frac{\sin[\bar \theta\bar\zeta(\bar z)]}
{\sqrt{\bar \theta\,\bar\zeta'(\bar z)}}~,
\\  \nonumber
\chi(z)=\frac{\sin[\theta\zeta(z)]}{\sqrt{\theta\,\zeta'(z)}}~,~~~~~~~~~
\bar\chi(\bar z)=\frac{\cos[\bar\theta\bar\zeta(\bar z)]}
{\sqrt{\bar \theta\,\bar\zeta'(\bar z)}}~,
\end{eqnarray}
where $\zeta(z)$ and $\zeta(\bar z)$ are monotonic functions with
the monodromies (\ref{monodromy of zeta}) and $\theta$ and
$\bar\theta$ are positive numbers, which we specify below. The
monodromies of these functions
\begin{eqnarray}\label{monodromy matrix}
M=\left( \begin{array}{cr}~~\cos(2\pi \theta)& \sin(2\pi \theta)\\
-\sin(2\pi \theta)&\cos(2\pi \theta)\end{array}\right)~,
~~~~~~~\bar M=\left( \begin{array}{cr}~~\cos(2\pi\bar\theta)&
\sin(2\pi\bar\theta)\\
-\sin(2\pi\bar\theta)&\cos(2\pi\bar\theta)\end{array}\right)
\end{eqnarray}
belong to the elliptic class and the periodicity  requires $\bar\theta
=\theta+m$, with an integer $m$.

Like in (\ref{psi mapsto psi}), eq. (\ref{elliptic solutions}) can
be considered as a Virasoro orbit of trigonometric functions, which
are solutions of the Hill equations for $T(z)=-\theta^2$ and $\bar
T(\bar z)=-\bar\theta^2$. The chiral stress tensor obtained from
(\ref{elliptic solutions})
\begin{equation}\label{transf of T}
T(z)=\frac{\psi''(z)}{\psi(z)}= -\theta^2\zeta'^2(z)+
\frac{1}{4}\left(\frac{\zeta''(z)}{\zeta'(z)}\right)^2
-\frac{1}{2}\left(\frac{\zeta''(z)}{\zeta'(z)}\right)'~,
\end{equation}
corresponds to the transformation law of $T(z)$ with the Schwarz
derivative. Then, the parameter $\theta$ defined by
\begin{equation}\label{a=}
\theta^2\int_0^{2\pi}dz\,\zeta'^2(z)=
\frac{1}{4}\int_0^{2\pi}dz\,\left(\frac{\zeta''(z)}{\zeta'(z)}\right)^2~,
\end{equation}
provides vanishing of the energy functional (\ref{E=0}).

The induced metric (\ref{induced metric(5)}) calculated from
(\ref{elliptic solutions}) reads
\begin{equation}\label{induced metric(6)}
e^\alpha=\theta\bar\theta\,\zeta'(z)\bar\zeta'(\bar z) \Big(A(z,\bar
z)\,\sin[\theta\zeta(z)+\bar\theta\bar\zeta(\bar z)]+B(z,\bar
z)\,\cos[\theta\zeta(z)+\bar\theta\bar\zeta(\bar z)]\Big)^2~,
\end{equation}
where
\begin{equation}\label{A,B}
A(z,\bar z)=1-\frac{\zeta''(z)\,\bar\zeta''(\bar z)}{4\theta\bar
\theta\,\zeta'\,^2(z)\,\,\bar\zeta'\,^2(\bar z)}~,~~~~~ B(z,\bar z)=
\frac{\zeta''(z)}{2\theta\,\zeta'\,^2(z)}+\frac{\bar\zeta''(\bar
z)}{2\bar\theta\,\bar\zeta'\,^2(\bar z)}~.
\end{equation}
With the parameterization  $\,A=1-\tan
\gamma(z)\,\tan\bar\gamma(\bar z)\,$ and $\,B=\tan
\gamma(z)+\tan\bar\gamma(\bar z),\,$ the conformal factor
(\ref{induced metric(6)}) takes the form
\begin{equation}\label{induced metric(6')}
e^\alpha=\theta\,\bar\theta\,\zeta'(z)\,\bar\zeta'(\bar z)\,R^2\,
\sin^2[\theta\,\zeta(z)+\bar\theta\,\bar\zeta(\bar
z)+\gamma(z)+\bar\gamma(\bar z)]~,
\end{equation}
where $R^2=A^2+B^2=[1+\tan^2\gamma(z)][1+\tan^2\bar\gamma(\bar z)].$
The solutions of the equation
\begin{equation}\label{sig. points}
\theta\,\zeta(z)+\bar\theta\,\bar\zeta(z)+\gamma(z)+\bar\gamma(\bar
z)=n\pi~,
\end{equation}
define the worldsheet singular points. Since the chiral and
antichiral parts of $B(z,\bar z)$ are periodic functions, the
introduced angle variables are bounded by
$-\frac{\pi}{2}<\gamma(z)<\frac{\pi}{2}$. 
The functions $\zeta(z)$ and $\bar\zeta(z)$, in contrast, are unbounded.
Hence, eq. (\ref{sig. points}) always has solutions and 
the string surfaces are singular.

\subsection{Liouville gauge for $SL(2,\rr)$ string}

Now we consider the Liouville gauge (\ref{Liouville gauge})
and the  surfaces related to the ground state functions
(\ref{ground state}).
As in the previous subsection, we use the correspondence between
the worldsheet tangent vectors in $\rr^{1,2}$
and the $SL(2,\rr)$ Kac-Moody currents.

The tangent vectors (\ref{B=sin-cos}) are equivalent to the currents
\begin{eqnarray}\label{J=sin-cos}
J(z)=\Lambda\,[t_0+\cos(nz)\,t_1+ \sin(nz)\,t_2]~,~~~
\bar J(\bar z)=\bar\Lambda\,[t_0-\cos(\bar n\bar z)\,t_1+\sin(\bar
n\bar z)\,t_2]\,.~~~
\end{eqnarray}

Note that the periodicity condition here does not require
necessarily $\Lambda=\bar\Lambda$. However, if
$\Lambda=\bar\Lambda$, the currents (\ref{J=sin-cos}) have equal
constant $t_0$ components $J_0=\Lambda=\bar J_0.$ These constraints
correspond to the vector gauged $SL(2,\rr)/U(1)$ coset model
\cite{Goddard:1984vk}. A more well investigated case is the axial gauged
$SL(2,\rr)/U(1)$ model \cite{W2}, which corresponds to
$J_0=\Lambda=-\bar J_0.$ This model was integrated in \cite{MW}
similarly to Liouville theory, and in the periodic case a free-field
parameterization was obtained in the hyperbolic sector. 
The exact integrability of the model
has been generalized  in \cite{Nikolai} to the
vector gauged model and to the elliptic sector of both models as
well. We use the corresponding technique to integrate the Kac-Moody
currents (\ref{J=sin-cos}) to WZW-fields.

\vspace{3mm}

The currents (\ref{J=sin-cos}) can be written in the form (see
(\ref{e^-t e^t}))
\begin{eqnarray}\label{J=t0+t1}
J(z)=\Lambda\,e^{\frac{1}{2}\,n
z\,t_0}\,(t_0+t_1)\,e^{-\frac{1}{2}\, n z\,t_0}~, ~~~~~~~~~ \bar
J(\bar z)=\bar\Lambda\,e^{-\frac{1}{2}\,\bar n\bar z\,t_0}
\,(t_0-t_1)\,e^{\frac{1}{2}\,\bar n\bar z\,t_0}~,
\end{eqnarray}
and with $h_l(z)=e^{-\frac{1}{2}\, n z\,t_0}\,g_l(z)\,$ and
$\,h_r(\bar z)=g_r(\bar z)\,e^{-\frac{1}{2}\,\bar n\bar z\,t_0},$
eq. (\ref{g'=jg}) reduces to
\begin{equation}\label{h'=jh}
h_l'(z)= \left[\left(\Lambda-\frac{n}{2}\right)t_0+ \Lambda
t_1\right]h_l(z)~,~~~~~~~~ h_r'(\bar z)= h_r(\bar
z)\left[\left(\bar\Lambda-\frac{\bar n}{2} \right)t_0-\bar\Lambda
t_1\right]~.
\end{equation}
The integration is then straightforward and leads to
\begin{equation}\label{LG g=}
g(z,\bar z)=e^{\frac{1}{2}\, n z\,t_0}\, e^{\frac{1}{2}\,
z\,a}\,g_{_0}\,e^{\frac{ 1}{2}\,\bar z\,\bar a}\,
e^{\frac{1}{2}\,\bar n \bar z\,t_0}~,
\end{equation}
where $g_{_0}$ is a $SL(2,\rr)$ valued integration constant and
\begin{equation}\label{A,bar A}
a= \left(2\Lambda- n\right)t_0+2\Lambda\, t_1~,~~~~~~~~\bar a =
\left(2\bar\Lambda-\bar n \right)t_0-2\bar\Lambda\, t_1~.
\end{equation}
The periodicity of (\ref{LG g=}) imposes the following condition
\begin{equation}\label{periodicity of g}
g_{_0}^{-1}\,e^{\pi a}\,g_{_0}=(-)^{n-\bar n}\, e^{\pi\bar a}~.
\end{equation}

Equations (\ref{LG g=})-(\ref{periodicity of g}) define
the WZW-fields $g(z,\bar z)$ and, thereby, the string surfaces in $SL(2,\rr)$.
However, to describe the structure of these surfaces in detail, similarly
to the flat case, an additional labour is needed.

Let us consider timelike $a$ and $\bar a$. In this case (see
(\ref{e^A})) the exponents in (\ref{periodicity of g}) take the form
\begin{equation}\label{e^Az}
e^{\pi\,a}=\cos\left(\pi\theta \right)\,I\,+
\sin\left(\pi\theta\right)\,\hat a~,~~~~~
e^{\pi\,\bar a}=\cos\left(\pi\bar\theta\right)\,I\,+
\sin\left(\pi\bar\theta\right)\,\hat{\bar a}~,
\end{equation}
where
\begin{equation}\label{theta=}
\theta=\sqrt{|\langle\,a\,a\,\rangle|}~,~~~~~~~~~~~~
\bar\theta=\sqrt{|\langle\,\bar a\,\bar a\,\rangle|}~,
\end{equation}
and $\hat a,\,$  $\,\hat{\bar a}$ are the normalized $sl(2,\rr)$
matrixes
\begin{equation}\label{hat a}
\hat a=\frac{a}{\theta}= \left( \begin{array}{cr}
0&-\frac{\theta}{n}\\ \frac{n}{\theta}&0~
\end{array}\right)~,~~~~~~
\hat{\bar a}=\frac{\bar a}{\bar\theta}= \left( \begin{array}{cr}
~0&-\frac{\bar n}{\bar\theta}\\ \frac{\bar\theta}{\bar n}&0
\end{array}\right)~.
\end{equation}
From (\ref{periodicity of g}) and (\ref{e^Az}) follows the equation
for $g_{_0}$
\begin{equation}\label{ag_0=}
g_{_0}^{-1}\,\hat a\,g_{_0}=s\,\hat{\bar a}~,
\end{equation}
and the relations between $\theta$ and $\bar\theta$
\begin{equation}\label{theta=bar theta}
\cos(\pi\theta)=\epsilon\cos(\pi\bar\theta)~,~~~~~~~~
\sin(\pi\theta)=s\,\epsilon\sin(\pi\bar\theta)~.
\end{equation}
Here $\epsilon=(-)^{n-\bar n}$ and $s=\pm\, 1,$ since the similarity
transformation $\hat a\mapsto g_{_0}^{-1}\,\hat a\,g_{_0}$ preserves
the norm of $\hat a$. Another invariant of the transformation $\hat
a\mapsto g_{_0}^{-1}\,\hat a\,g_{_0}$ is the sign of $\hat a^0$,
where $\hat a^0$ is the $t_0$ component of $\hat a$. The $t_0$
components of the unit vectors (\ref{hat a}) are
\begin{equation}\label{t_0 components}
\hat a^0=-\frac{\theta^2+n^2}{2n\theta}~,~~~~~~~~~ \hat{\bar
a}^0=-\frac{\bar\theta^2+\bar n^2}{2\bar n\bar\theta}~,
\end{equation}
and we conclude that $\hat\epsilon=\mbox{sign}(n\,\bar n)$.

The solution of (\ref{ag_0=}) is given by
\begin{equation}\label{g_0(+)}
g_{_0}=\frac{1}{\sqrt{s\, n\,\bar n\,\theta\bar\theta}}\left(
\begin{array}{cr} ~~\theta\,\bar\theta\,\cos\phi&~~
-\theta\,\bar n\,\sin\phi \\ \\
s\,n\,\bar\theta\,\sin\phi&  s\,n\,\bar
n\,\cos\phi\end{array}\right)~,
\end{equation}
where $\phi$ is an angle variable, which parameterize the freedom in
$g_{_0}$. In addition, eqs. (\ref{ag_0=}) and (\ref{LG g=}) yield
\begin{equation}\label{e(za)}
g(z,\bar z)=e^{\frac{1}{2}\, n z\,t_0}\, e^{\frac{1}{2}\,(\theta
z+s\,\bar\theta\bar z)\,\hat a}\,g_{_0}\, e^{\frac{1}{2}\,\bar n
\bar z\,t_0}~.
\end{equation}

To visualize the constructed worldsheets (\ref{e(za)}), we introduce
two complex planes defined by the embedding coordinates (\ref{Y=g})
\begin{equation}\label{Z=}
Z=Y_1+iY_2=\langle(t_1+it_2)\,g\,\rangle~,~~~~~Z^0=Y^{\tilde0}+iY^0=
\langle(I-it_0)\,g\,\rangle~.
\end{equation}
Taking into account that $t_0$ is the generator of rotations in
$(t_1, t_2)$ and $(I,t_0)$ planes
(see (\ref{(t1+it2)e^t})-(\ref{(I+it0)e^t})), from (\ref{e(za)})
we obtain
\begin{eqnarray}\nonumber
Z=e^{\frac{i}{2}\,(nz-\bar n\bar z)}\, \left[A_+\,\,
e^{\frac{i}{2}\,(\theta z+s\,\bar\theta\bar z)} +A_-\,\,
e^{-\frac{i}{2}\,(\theta z+s\,\bar\theta\bar z)} \right]~, \\
\label{Z,Z^0}\\
\nonumber Z^0=e^{\frac{i}{2}\,(nz+\bar n\bar z)}\, \left[B_+\,\,
e^{\frac{i}{2}\,(\theta z+s\,\bar\theta\bar z)} +\,B_-\,\,
e^{-\frac{i}{2}\,(\theta z+s\,\bar\theta\bar z)} \right]~.
\end{eqnarray}
The coefficients $A_\pm$ and $B_\pm$ correspond to the following
normalized traces
\begin{eqnarray}\nonumber
A_\pm=\frac{1}{2}\,\langle\,(t_1+it_2)\,(I\,\mp\, i\hat
a)\,g_{_{0}}\,\rangle =i\,\frac{s\,(\theta\,\mp\,
n)(s\,\bar\theta\,\pm\,\bar
n)\,e^{\pm\,is\,\phi}}{4\sqrt{s\,n\bar n\,\theta\bar\theta}}~,\\
\label{Bpm}
\\ \nonumber
B_\pm=\frac{1}{2}\,\langle\,(I-it_0)\,(I\,\mp\, i\hat
a)\,g_{_{0}}\,\rangle=\, ~\frac{s\,(\theta\,\mp\,
n)(s\,\bar\theta\,\mp\,\bar n)\,e^{\pm\,is\,\phi}} {4\sqrt{s\,n\bar
n\,\theta\bar\theta}}~.~
\end{eqnarray}
The conformal factor of the induced metric tensor obtained from
these equations reads
\begin{equation}\label{e^alpha AdS 3}
e^{\alpha}=\mbox{Re}\left(\partial Z_0\,\bar\partial Z_0^*-
\partial Z\,\bar\partial Z^*\right)
=\frac{s\,(\theta^2-n^2)(\bar\theta^2-\bar
n^2)}{16\theta\bar\theta}\, [1+\cos(\theta z+s\,\bar\theta\bar
z+2s\,\phi)]~.
\end{equation}

According to (\ref{theta=}) and (\ref{A,bar A}), the parameters of the
solutions are related by 
\begin{equation}\label{theta=n,Lambda}
\theta^2=n^2-4\Lambda\,n~,~~~~~~~\bar\theta^2= \bar
n^2-4\bar\Lambda\,\bar n~.
\end{equation}
These relations reduce the conformal factor (\ref{e^alpha AdS 3}) to
\begin{equation}\label{e^alpha AdS 3=}
e^{\alpha}=\frac{\Lambda\,\bar\Lambda\,|n\,\bar
n|}{\theta\bar\theta}\, [1+\cos(\theta z+s\,\bar\theta\bar
z+2s\,\phi)]~.
\end{equation}
Since $\Lambda,$ $\bar\Lambda,$ $\theta,$ $\bar\theta$ are positive,
the metric is regular almost everywhere. The singular points
correspond to the solutions of the equation $1+\cos(\theta
z+s\,\bar\theta\bar z+2s\,\phi)=0,$ i.e.
\begin{equation}\label{zeros of e^alpha 1}
(\theta+s\,\bar\theta)\tau+(\theta-s\,\bar\theta)\sigma=(2m+1)\pi
-2s\,\phi~ ~~~~~(m\in \zz)~.
\end{equation}
Note that eqs. (\ref{e^alpha AdS 3=})-(\ref{zeros of e^alpha 1}) are
similar to (\ref{e^alpha=})-(\ref{zeros of e^alpha}), derived for
the spiky and circular strings in $\rr^{1,2}$.

\vspace{3mm}

Now we use the relation (\ref{theta=bar theta}) to exclude one
continuous parameter.  The case $\epsilon=1$ corresponds to even
$n\pm\bar n$. According to (\ref{theta=bar theta}),
$\theta-s\,\bar\theta$  is also even for $\epsilon=1$. With the
notations
\begin{equation}\label{omega, even}
\theta-s\,\bar\theta=2\nu~,~~~~~\theta+s\,\bar\theta=2\mu~,
~~~~~n-\bar n=2k~,~~~~~n+\bar n=2l~,
\end{equation}
the solution (\ref{Z,Z^0}) can be written as
\begin{eqnarray}\nonumber
Z(\tau,\sigma)=e^{i(k\tau+l\sigma)}\, \left[A_+\,\,
e^{i(\mu\tau+\nu\sigma)} +A_-\,\,
e^{-i(\mu\tau+\nu\sigma)} \right]~,~ \\
\label{Z,epsilon=1}\\
\nonumber Z^0(\tau,\sigma)=e^{i(l\tau+k\sigma)}\, \left[B_+\,\,
e^{i(\mu\tau+\nu\sigma)} +\,B_-\,\, e^{-i(\mu\tau+\nu\sigma)}
\right]~.
\end{eqnarray}
The parameters ($\nu$, $k$, $l$) here are integer, which provides
the periodicity of (\ref{Z,epsilon=1}).

If $\epsilon=-1$, $n\pm\bar n$ and $\theta-s\,\bar\theta$ are odd.
Therefore, instead of (\ref{omega, even}) we use the notations
\begin{equation}\label{omega, odd}
\theta-s\,\bar\theta=2\nu+1~,~~~~~\theta+s\,\bar\theta=2\mu+1~,
~~~~~n-\bar n=2k+1~,~~~~~n+\bar n=2l+1~,
\end{equation}
again with integer $\nu$, $k$ and $l$. The solution (\ref{Z,Z^0}) in
this case becomes
\begin{eqnarray}\nonumber
Z(\tau,\sigma)=e^{i(k\tau+l\sigma)}\, \left[A_+\,\,
e^{i[(\mu+1)\tau+(\nu+1)\sigma]} +A_-\,\,
e^{-i(\mu\tau+\nu\sigma)} \right]~,~ \\
\label{Z,epsilon=-1}\\
\nonumber Z^0(\tau,\sigma)=e^{i(l\tau+k\sigma)}\, \left[B_+\,\,
e^{i[(\mu+1)\tau+(\nu+1)\sigma]} +\,B_-\,\,
e^{-i(\mu\tau+\nu\sigma)} \right]~.
\end{eqnarray}

One has to remember that the parameters of the solutions are
restricted by
\begin{equation}\label{conditions}
n\neq 0~,~~~~\bar n\neq 0~;~~~~~\theta>0~,~~~~~\bar\theta>0;
~~~~~\theta\neq |n|~,~~~~\bar\theta\neq |\bar n|~,
\end{equation}
where the last two inequalities follow from (\ref{theta=n,Lambda}). If
these conditions are not fulfilled, the coefficients $A_\pm$ and
$B_\pm$ are either singular, or vanishing. The vanishing of the
coefficients correspond to the degenerated case
$\Lambda=0=\bar\Lambda$. The formulation of the conditions
(\ref{conditions}) in terms of the new parameters ($\mu$, $\nu$,
$k$, $l$) is more complicated. Therefore, sometimes it is more
convenient to keep the old parameters ($n$, $\bar n$, $\theta$,
$\bar\theta$).

The spatial part of $AdS_3$ is given by the complex plane $Z$. Due
to the similarity with the flat space, the function $Z$ in
(\ref{Z,Z^0}) can be described in a same manner as
(\ref{Z(tau,sigma) in R}) for $\rr^{1,2}$ strings. The case
$\theta-s\,\bar\theta=0$ is special, like $n-\bar n=0$ in
$\rr^{1,2}$, and we consider it here separately.

The condition $\theta-s\,\bar\theta=0$ implies $s=1$ and
$\theta=\bar\theta$. The corresponding solutions of eq. (\ref{zeros
of e^alpha 1}) are $\sigma$-independent discrete values of $\tau$,
like for the circular oscillating strings in $\rr^{1,2}$. For
simplicity, let's assume $n=\bar n$ and $2\phi=-\pi$. The solution
(\ref{Z,Z^0})-(\ref{Bpm}) then reduces to
\begin{eqnarray}\label{Z=3-4}
Z=\frac{n^2-\theta^2}{2i\,n\,\theta}\,\, \sin
(\theta\tau)\,e^{in\sigma}~,~~~~~~
 Z^0=\left(\frac{n^2+\theta^2}{2n\,\theta}\,
\sin(\theta\tau)+i\cos(\theta\tau)\right)\,e^{in\tau}~,
\end{eqnarray}
and the conformal factor (\ref{e^alpha AdS 3}) becomes
\begin{equation}\label{e^alpha AdS 3-4}
e^\alpha =\frac{(n^2-\theta^2)^2}{8\theta^2}\, \sin^2(\theta\tau)~.
\end{equation}
The time variable in $AdS_3$ is given by the phase of $Z^0$, which
in (\ref{Z=3-4}) is only $\tau$ dependent. For a fixed $\tau$, the
function $Z$ in (\ref{Z=3-4}) provides a circle with the radius
proportional to $\sin (\theta\tau)$. Therefore, eq. (\ref{e^alpha
AdS 3-4}) describes a circular oscillating string like (\ref{X=()+})
in $\rr^{1,2}$. This type of string solutions were obtained earlier
in \cite{Larsen:1998sq},  where the authors
used the Pohlmeyer type scheme for the embedding space $\rr^{2,2}$, and
provided some non periodic solutions as well.

In the limit $\theta\rightarrow 0$  eqs. (\ref{Z=3-4})-(\ref{e^alpha
AdS 3-4}) are reduced to
\begin{equation}\label{Z=3p}
Z=\frac{n\,\tau}{2i}\,\,e^{in\sigma}~,~~~~~~~~
Z^0=\frac{1}{2}\,(n\tau+2i)\,e^{in\tau}~;~~~~~~e^\alpha
=\frac{n^4}{8}\,\tau^2~.
\end{equation}
This case correspond to a nilpotent $a$ in (\ref{hat a}), and the
related WZW-field belongs to the parabolic monodromy.

Eqs. (\ref{Z=3-4})-(\ref{e^alpha AdS 3-4}) allow a continuation to
imaginary $\theta$, which leads to the hyperbolic solutions
with
\begin{eqnarray}\label{Z=3}
Z=\frac{n^2+\theta^2}{2i\,n\theta}
\,\,\sinh(\theta\tau)\,\,e^{in\sigma}~,~~~~~
Z^0=\left(\frac{n^2+\theta^2}{2n\,\theta}\,
\sinh(\theta\tau)+i\cosh(\theta\tau)\right)\,e^{in\tau}~,
\end{eqnarray}
and the conformal factor
\begin{equation}\label{e^alpha AdS}
e^\alpha =\frac{(\theta^2+n^2)^2}{8\theta^2}\, \,
\sinh^2(\theta\tau)~.
\end{equation}
These parabolic and hyperbolic solutions shrink to the origin $Z=0$ 
only once at $\tau=0$. One
can show that other parabolic and hyperbolic solutions corresponding
to $n\neq \bar n$  are also obtained
by the analytical continuation of the elliptic  solutions
(\ref{Z,Z^0}) with $s=1$ and $\theta =\bar\theta$.

\vspace{3mm}

Let's assume now that $\theta-s\,\bar\theta\neq 0$. The function
$Z(\tau,\sigma)$ in (\ref{Z,Z^0}) then fulfills the relation
(\ref{X1-X_2=1}) with
\begin{equation}\label{AdS omega, omega0}
\omega=-\frac{\theta\,\bar n+n\,s\,
\bar\theta}{\theta-s\,\bar\theta}~,~~~~~~~~~
\omega_0=\frac{\theta+s\,\bar\theta}{\theta-s\,\bar\theta}~,
\end{equation}
and the initial configuration
\begin{equation}\label{AdS Z(sigma)}
Z(\sigma)= \left[A_+\,\, e^{\frac{i}{2}\,(n+\bar n+\theta
-s\,\bar\theta)\sigma} +A_-\,\,e^{\frac{i}{2}\,(n+\bar n-\theta
z+s\,\bar\theta\bar)\sigma}\right]~.
\end{equation}
As in the flat case, the dynamics in $\tau$ preserves the shape of
this closed curve. Effectively it rotates only. The curve defined by
(\ref{AdS Z(sigma)}) is represented again as a combination of two
rotations. Therefore, the properties of the spiky strings, discussed
in Subsection 2.3 and Appendix E, can be generalized to this case.

\vspace{3mm}

Finally, we briefly describe the general case with the Kac-Moody 
current
\begin{eqnarray}\label{J general}
J(z)=\frac{\Lambda}{\zeta'(z)}\,
[t_0+\cos(n\zeta(z))\,t_1+ \sin(n\zeta(z))\,t_2]~,~~~
\end{eqnarray}
which corresponds to the tangent vector (\ref{B=()1}). The description
of the antichiral part is similar. Using the same trick as in 
(\ref{J=t0+t1}), eq. (\ref{g'=jg}) reduces to
\begin{equation}\label{h'=jh general}
h_l'(z)= \left[\frac{\Lambda}{\zeta'(z)}\,
(t_0+ t_1)-\frac{n\,\zeta'(z)}{2}\,t_0\right]\,h_l(z)~,
\end{equation}
with $h_l(z)=e^{-\frac{1}{2}\, n\zeta(z)\,t_0}\,g_l(z).$ 
The change of variable $z \mapsto\zeta$ in (\ref{h'=jh general}) yields
\begin{equation}\label{h'=jh general 1}
\partial_\zeta h_l(\zeta)= \left[\Lambda \rho^2(\zeta)\,
(t_0+ t_1)-\frac{n}{2}\,t_0\right]\,h_l(\zeta)~,
\end{equation}
where $\rho(\zeta)=\left(\frac{dz}{d\zeta}\right)^2.$ The chiral current,
which stands in this equation has vanishing $t_2$ and constant $t_-$
components. Therefore, it can be interpreted 
as a current of nilpotently gauged WZW theory in the gauge 
$J_2=0$. This reduction is described again by Liouville theory and one can use 
the scheme of the previous subsection.

\section{Summary}

Here we give a summary of the paper and describe some unsolved
problems. 
In summary we list the following items:

1. On the basis of the isometry between $\rr^{1,2}$ and $sl(2,\rr)$,
the Pohlmeyer scheme for string dynamics in $\rr^{1,2}$ and
$SL(2,\rr)$ is formulated in equivalent forms. This equivalence maps
the tangent vectors of the $\rr^{1,2}$ string surfaces to the
$SL(2,\rr)$ Kac-Moody currents, which obey the Virasoro constraints.

2. The closed string dynamics in $\rr^{1,2}$ is integrated within
the Pohlmeyer scheme, using the parameterization (\ref{u=f'}) for
the components of the fundamental quadratic forms. These
parameterization fixes the conformal gauge freedom up to
translations and a one parameter subgroup. The factorization of the
set of solutions by the remaining conformal freedom provides the string
surfaces in the lightcone gauge. These surfaces have 
a  degenerated induced metric and the chiral
components of the second fundamental form $u(z),$
$\bar u(\bar z)$ do not have fixed signs in the interval of
periodicity.

3. The second class of closed string surfaces in $\rr^{1,2}$
corresponds to the case with non vanishing $u(z)$ and $\bar u(\bar
z)$. In the gauge where $u(z)$ and $\bar u(\bar
z)$ are constant, the Gau\ss $~$ equation  reduces to the Liouville
equation and, on the basis of its general solution, the linear system
of the Pohlmeyer scheme is integrated in the form (\ref{B=F}) and
(\ref{B=psi}). The periodicity condition is satisfied by the class
of singular Liouville fields with the monodromy matrix $\pm I$. This
class is parameterized by the Virasoro coadjoint orbits, which
are labeled by two nonzero integers ($n$, $\bar n$).
The signs of $n$ and $\bar n$ coincide with the signs of $u$ and
$\bar u$, respectively.
The vacuum Liouville fields with constant stress tensor
$T(z)=-\frac{n^2}{4}$, $\bar T(\bar z)=-\frac{\bar n^2}{4}$ describe
oscillating circular  (if $n=\bar n$) and rotating spiky
(if $n\neq\bar n$) strings.  
The number of spikes is equal to $|n-\bar n|$ and
the shape of string configurations at a fixed time  essentially
depends on the sign of $n\bar n$.

4. The tangent vectors of $\rr^{1,2}$
string surfaces in the lightcone gauge correspond to 
Kac-Moody currents of a nilpotently gauged $SL(2,\rr)$ WZW model.
The related Liouville fields are singular
and have vanishing chiral energy functional. 
The corresponding string surfaces are described 
by a pair of monotonic functions $\zeta(z)$, $\,\bar\zeta(\bar z)$,
used in parameterization of 2d conformal group. 
These $SL(2,\rr)$  string surfaces are always singular, 
like the surfaces in $\rr^{1,2}$.

5. In the Liouville gauge, the vacuum  field configurations  
provide the coset Kac-Moody currents of the $SL(2,\rr)/U(1)$ model.
They are labeled by two nonzero integers $n$ and $\bar n$.
The integration of the currents in the elliptic sector of WZW-fields
leads to circular and spiky strings. 
These string surfaces split in four classes,
depending on the sign of $n\bar n$ and the parity of ${n-\bar n}$.  
The parabolic and hyperbolic solutions are obtained by the analytical
continuation of the elliptic solutions with $n\bar n>0$
and even $n-\bar n$.

\vspace{3mm} 

The main result of the paper is the description of the new classes
of closed string solutions in $\rr^{1,2}$ and $SL(2,\rr)$.
They are given by  (\ref{B=()1}) in  $\rr^{1,2}$, and by
(\ref{nilpotent solutions}), (\ref{elliptic solutions}) and
 (\ref{Z,Z^0})-(\ref{Bpm}) in $SL(2,\rr)$.

\vspace{3mm} 

The construction of quantum theory of these solutions requires quantization of
singular Liouville fields. 
The quantized Liouville theory on a cylindrical spacetime 
\cite{BCT} and on its Euclidean
counterpart \cite{DO} are one of the most remarkable results
in 2d QFT. However, this theory
corresponds to the quantization of regular Liouville fields, 
which allow a free-field 
parameterization. This parameterization
is a basis for canonical quantization in the Minkowskian spacetime,
where the parameterizing field can be chosen as the $in$ (or $out$) field
of the theory. It also helps to construct the vertex operators 
\cite{Teschner:2003en} and calculate their causal algebra \cite{FJ}.
 
The singular Liouville fields, we are interested in, have 
a regular stress tensor. 
A natural way for a generalization of the canonical scheme to the singular case
is to find a free-field parameterization of these singular fields. 
Note that such Liouville fields on a plane allow 
a free-field parameterization. Namely,
a Liouville field with $N$ lines of singularities can be canonically
parameterized by one regular free field and the asymptotic data of
$N$ relativistic particles \cite{J}.  Unfortunately, a generalization 
of this result to the periodic case is still unknown
(see however \cite{Marn}).

The vacuum Liouville field configurations, used in the spiky strings,
for $n=1$ correspond to M\"obius invariant ground state, which arises
in boundary Liouville theory on a strip with Dirichlet conditions.
The Hamiltonian description of such field configurations was
considered in \cite{DJ}, and was an attempt 
to understand Liouville theory
with Dirichlet boundary conditions as a limit of the theory with
Neumann conditions \cite{DJ1}. 
The latter correspond to the elliptic monodromy \cite{GN}
and its Euclidean version is given by FZZT branes
\cite{FZZ}. This programme also needs further investigation. A
possible candidate for quantum theory of the spiky strings in
$\rr^{1,2}$ could be the `Wick
rotated' ZZ branes \cite{ZZ}.

Another possible approach to quantum theory of singular
Liouville fields is the geometric quantization 
on the Virasoro coadjoint orbits \cite{Alekseev:1988ce}. These orbits
parameterize nilpotently gauged strings in $SL(2,\rr)$ and spiky 
strings both in $\rr^{1,2}$ and $SL(2,\rr)$.

There is a renewed interest to Liouville theory caused by \cite{Alday:2009aq},
and it is a challenge to understand whether the singular Liouville fields
and spiky strings described in the present paper have any relation to it. 

 \vspace{5mm}

\noindent
{\bf Acknowledgment}\\[2mm]
The present paper is a continuation of the previous work done in
collaboration with Harald Dorn. I would like to thank Harald for
many illuminating discussions.
\\[1mm]
I also acknowledge very useful conversations with Nadav Drukker, Jan
Plefka and Donovan Young.
\\[1mm]
This work has been supported in part by Deutsche Forschungsgemeinschaft via
SFB 647 and also by GNSF.

\newpage

\setcounter{equation}{0}

\def\theequation{A.\arabic{equation}}

\noindent {\bf \large Appendix A}

\vspace{3mm}

\noindent In this appendix we integrate the linear system
(\ref{partial B}) with $u$, $\bar u$ and $e^\alpha$ given by
(\ref{u=f'}). We then, briefly consider the case of the Liouville
gauge, discussed in subsection 2.3.

Starting with the last equations in (\ref{partial B})
\begin{equation}\label{N=?}
\partial N=\frac{ 2f'(z)}{\big[\bar f(\bar z)-f(z)\big]^2}\,\bar
B(\bar z)~,~~~~~~~\bar\partial N= -\frac{ 2\bar f'(\bar
z)}{\big[\bar f(\bar z)-f(z)\big]^2} \,B(z)~,
\end{equation}
one obtains
\begin{equation}\label{N=F}
N=\frac{2\,\bar B(\bar z)}{\bar f(\bar z)-f(z)} +\bar b(\bar z)=
\frac{2\,B(z)}{\bar f(\bar z)-f(z)}+b(z)~,
\end{equation}
where $b(z)$ and $\bar b(\bar z)$ are chiral and antichiral fields
with values in $\rr^{1,2}$. Multiplying (\ref{N=F}) by the
denominator $\bar f(\bar z)-f(z)$ and then differentiating it by
$\bar\partial$ and $\partial$, one gets
\begin{equation}\label{b=}
b(z)=c\,f(z)+c_1~,~~~~~~~~\bar b(\bar z)=-c\,\bar f(\bar z)+\bar
c_1~,
\end{equation}
with $\rr^{1,2}$-valued constant vectors $c$, $c_1$, $\bar c_1$. The
insertion of (\ref{N=F})-(\ref{b=}) into the first equation of
(\ref{partial B}) leads to
\begin{equation}\label{B'=}
B'(z)=c_1\,f'(z)+c\,f(z)\,f'(z)~,
\end{equation}
which is integrated to
\begin{equation}\label{B=}
B(z)=c_1\,f(z)+\frac{c}{2}\,f^2(z)+ c_2~,
\end{equation}
where $c_2$ is a new $\rr^{1,2}$-valued integration constant. The
antichiral sector similarly yields
\begin{equation}\label{bar B=}
\bar B(\bar z)=-\bar c_1\,\bar f(\bar z)+\frac{c}{2}\,\bar f^2(\bar
z)+\bar c_2~.
\end{equation}
With these $B(z)$ and $\bar B(\bar z)\,$ eq. (\ref{N=F}) relates the
integration constants in the chiral and antichiral sectors by $\bar
c_1=-c_1,\,$  $\bar c_2=c_2.$ Then, (\ref{N=F}) reads
\begin{equation}\label{N=}
N(z,\bar z)=\frac{\bar f(\bar z)+f(z)}{\bar f(\bar
z)-f(z)}\,\,c_1+\frac{f(z)\,\bar f(\bar z)}{\bar f(\bar
z)-f(z)}\,\,c+\frac{2}{\bar f(\bar z)-f(z)}\,\,c_2~.
\end{equation}
Finally, using the notations $c_1\equiv {\bf e},$ $c\equiv 2{\bf
e}_{_-}$ and $c_2\equiv {\bf e}_{_+},\,$ one obtains $B$, $\bar B$
and $N$ in the form (\ref{B=f})-(\ref{N=f}).

\vspace{3mm} In the Liouville gauge, where $u$ and $\bar u$ are
constants and $e^\alpha$ is given by (\ref{Liouv-g-solution}), we
start again with the last equations in (\ref{partial B})
\begin{equation}\label{N=F,bar F}
\partial N=\frac{2\epsilon}{|\bar\lambda|}\,\,\frac{ F'(z)
 \,\bar F'(\bar z)}{\big[\epsilon\,F(z)+\bar\epsilon\,\bar F(\bar
z)\big]^2}\, \bar B(\bar z)~,~~~~~~~\bar\partial N=
\frac{2\bar\epsilon}{|\lambda|}\,\,\frac{F'(z)\,\bar F'(\bar
z)}{\big[\epsilon\,F(z)+\bar\epsilon\,\bar F(\bar z)\big]^2}
\,B(z)~.
\end{equation}
The integration steps here are as before. At the final stage one
gets the following relations for the integration constants $\bar
c_1=-c_1$, $\bar\epsilon\,\bar c_2=\epsilon\,c_2$,  and with the
notations $c_1\equiv {\bf e},\,$ $ c\equiv 2{\bf e}_{_-}\,$ and $\,
c_2\equiv \epsilon{\bf e}_{_+}$ one arrives at (\ref{B=F})-(\ref{N=F
bar F}).

\vspace{9mm}

\setcounter{equation}{0}

\def\theequation{B.\arabic{equation}}

\noindent {\bf \large Appendix B}

\vspace{3mm}

\noindent Here we analyze the freedom of conformal transformations
in the gauge (\ref{u=f'}). Let's write these transformations in the
infinitesimal form
\begin{equation}\label{small zeta}
\zeta(z)= z+\varepsilon\phi(z)~,~~~~~~ \bar\zeta(\bar z)=\bar
z+\varepsilon\bar\phi(\bar z)~,
\end{equation}
where $\phi(z)$ and $\bar\phi(\bar z)$ are periodic functions and
$\varepsilon$ is an infinitesimal parameter. Keeping the first order
terms in $\varepsilon$, from (\ref{alpha mapsto})-(\ref{u mapsto})
and (\ref{u=f'}) we find
\begin{eqnarray}\label{f+f mapsto}
(\bar f-f)^2\mapsto (\bar f-f)^2+ \varepsilon [(\bar f-f)^2
(\phi'+\bar\phi')+2(\bar f-f)(\bar f'\,\bar\phi-f'\,\phi)]~,
\\ \nonumber \\
\label{infin transf} u\mapsto u+\varepsilon(2\phi'\,f'+\phi\,f'')~,
~~~~~~~~~~~~\bar u \mapsto \bar u-\varepsilon(2\bar\phi'\,\bar f'+
\bar\phi\,\bar f'')~.
\end{eqnarray}
These transformations preserve the gauge (\ref{u=f'}), if they
correspond to infinitesimal shifts $f\mapsto f+\varepsilon\rho\,$
and $\bar f\mapsto\bar f+\varepsilon\bar\rho$, with some chiral
$\rho(z)$ and antichiral $\bar\rho(\bar z)$ functions. By (\ref{f+f
mapsto}) one then finds the relation
\begin{equation}\label{rho+bar rho}
2(\bar\rho-\rho)= (\bar f-f)(\phi'+\bar\phi')+2(\bar
f'\,\bar\phi-f'\,\phi)~.
\end{equation}
Its differentiation in $z$ and $\bar z$ leads to the equation $\bar
f'\,\phi''-f'\,\bar\phi''=0,$ which is solved by
\begin{equation}\label{phi''=f'}
\phi''(z)=c\,f'(z)~,~~~~~~~~\bar\phi''(\bar z)=c\,\bar f'(\bar z)~,
\end{equation}
with a constant $c$. The integration of this equation in two steps
provides
\begin{equation}\label{phi'=f}
\phi'(z)=c\,(f(z)-p)~,~~~~~~~~\bar\phi'(\bar z)=c\,(\bar f(\bar
z)-p)~,
\end{equation}
and
\begin{equation}\label{phi=}
\phi(z)=\phi_0+i\,c\sum_{n\neq 0} \frac{a_n}{n}\,e^{-inz}~,~~~~~~~~
\bar\phi(\bar z)=\bar\phi_0+i\,c\sum_{n\neq 0}\frac{\bar a_n}{n}
\,e^{-in\bar z}~.
\end{equation}
Here we have used the Fourier mode expansions (\ref{Fourier modes})
for $f(z)\,$ and  $\bar f(\bar z),$ the equality of their zero modes
(\ref{p=bar p}) and the periodicity of $\phi\,$ and $\bar\phi$.

From (\ref{phi'=f}) and (\ref{rho+bar rho}) we also find
\begin{equation}\label{rho, bar rho=}
\rho=\frac{1}{2c}\,\phi'^{\,2} +\frac{1}{c}\,\phi''\,\phi~,~~~~~~~
\bar\rho=\frac{1}{2c}\,\bar\phi'^{\,2}
+\frac{1}{c}\,\bar\phi''\,\bar\phi~.
\end{equation}
Note that $c=0$ corresponds to translations $\phi(z)=\phi_0,\,$
$\bar\phi(\bar z)=\bar\phi_0,$ with constant $\phi_0\,$ and
$\bar\phi_0.$ In this case (\ref{rho+bar rho}) provides
$\rho(z)=\phi_0\, f'(z)$ and $\bar\rho(\bar z)=\bar\phi_0\, \bar
f'(\bar z).$

Eqs. (\ref{rho, bar rho=}) and (\ref{phi''=f'}) allow to write
(\ref{infin transf}) in the form $u\mapsto u+\varepsilon\,\rho',$
$\bar u\mapsto \bar u+\varepsilon\,\bar\rho'.$ This form of
transformations corresponds to $f\mapsto f+\varepsilon\rho,\,$
$\,\bar f\mapsto\bar f+\varepsilon\bar\rho$, and proves that the
described conformal transformations preserve the gauge (\ref{u=f'}).

Finally we show that the conditions (\ref{p=bar p}) are invariant
under these conformal transformations. First recall that the zero
mode of a periodic function is given by the integral of this
function over the period. Since $f$ and $\bar f,$ as well as $f^2$
and $\bar f^2,$ have equal zero modes, from (\ref{phi'=f}) follows
that the zero modes of $\phi'^{\,2}$ and $\bar\phi\,'^{\,2}$ are
also equal. Then, due to (\ref{rho, bar rho=}), $\rho$ and $\bar\rho$
have equal zero modes too. This proves the invariance of the first
relation in (\ref{p=bar p}).

To prove the second relation, one has to compare the zero modes of
$f\,\rho$ and $\bar f\,\bar\rho$ (the first order terms in
$\varepsilon$). Here, it is convenient to use (\ref{phi'=f}),
(\ref{rho, bar rho=}) and express $f\,\rho$ through $\phi$ and its
derivatives. Then, a part of the zero mode of $f\,\rho$  vanishes
due to the relation
$\phi'^{\,3}+2\phi''\,\phi'\,\phi=(\phi'^{\,2}\,\phi)'$. The same is
valid for the zero mode of $\bar f\,\bar\rho$. The rest parts of the
integrals $\int_0^{2\pi} \mbox{d}z\,\,f(z)\rho(z)$ and
$\int_0^{2\pi} \mbox{d}\bar z\,\,\bar f(\bar z)\bar\rho(\bar z)$
coincide trivially.

\vspace{9mm}

\setcounter{equation}{0}

\def\theequation{C.\arabic{equation}}

\noindent {\bf \large Appendix C}

\vspace{3mm}

\noindent The exponential $V=e^{\frac{\alpha}{2}}$ has some
remarkable properties in Liouville theory. The conformal weight of
$V$ is $-\frac{1}{2}.$ The Liouville equation (\ref{Liouville eq})
is equivalent to the quadratic relation for $V$
\begin{equation}\label{Liouville eq V}
V\,\partial\bar\partial V-\partial V\bar\partial
V=-\frac{\lambda\bar\lambda}{2}~.
\end{equation}
In addition, $V$ fulfills the following linear relations
\begin{equation}\label{dV=TV}
\partial^2V(z,\bar z)=T(z)\,V(z,\bar z)~,~~~~~~~~
\bar\partial^2V(z,\bar z)=\bar T(\bar z)\,V(z,\bar z)~,
\end{equation}
where
\begin{equation}\label{T}
T=\frac{1}{4}\,(\partial\alpha)^2+\frac{1}{2}\,\partial^2\alpha~,
~~~~~\bar T= \frac{1}{4}\,(\bar\partial\alpha)^2+\frac{1}{2}\,
\bar\partial^2\alpha~,
\end{equation}
are chiral ($\bar\partial T=0$) and antichiral ($\partial\bar T=0$)
components of the stress tensor of Liouville theory.

The linear equation (\ref{dV=TV}) defines the representation
$V=A\,\big|\psi(z)\bar\psi(\bar z)+\chi(z)\bar\chi(\bar z)\big|,$
where $A$ is a normalization constant and the functions $\psi(z)$,
$\chi(z)$  and $\bar\psi(\bar z)$, $\bar\chi(\bar z)$ are solutions
of the chiral and antichiral Hill equations, respectively:
\begin{equation}\label{Hill eq}
\Psi''(z)=T(z)\,\Psi(z)~,~~~~~~~~\bar\Psi''(\bar z)=\bar T(\bar
z)\,\bar\Psi(\bar z)~.
\end{equation}
Choosing the Wronskians of these equations by
\begin{equation}\label{Wronskians}
\psi(z)\chi'(z)-\psi'(z)\chi(z)=\epsilon~,~~~~ \bar\psi(\bar
z)\bar\chi'(\bar z)-\bar\psi'(\bar z)\bar\chi(\bar z)=-\bar\epsilon
\end{equation}
($\epsilon=\mbox{sign}\lambda,$
$\,\bar\epsilon=\mbox{sign}\bar\lambda$), one can fix the
normalization constant $A$ from (\ref{Liouville eq V}), and obtain
\begin{equation}\label{V=}
e^{\frac{\alpha}{2}}=\sqrt{\frac{|\lambda\bar\lambda|}{2}}\,
\big|\psi(z)\bar\psi(\bar z)+\chi(z)\bar\chi(\bar z)\big|~.
\end{equation}
Introducing
\begin{equation}\label{F=}
F(z)=\epsilon\,\,\frac{\chi(z)}{\psi(z)}~,~~~~~~ \bar F(\bar
z)=\bar\epsilon\,\,\frac{\bar\psi(\bar z)}{\bar\chi(\bar z)}~,
\end{equation}
one finds from (\ref{Wronskians})
\begin{equation}\label{F'=}
F'(z)=\frac{1}{\psi^2(z)}~,~~~~~~ \bar F'(\bar
z)=\frac{1}{\bar\chi^2(\bar z)}~.
\end{equation}
Then, (\ref{V=}) takes the Liouville form of the general solution
(\ref{Liouv-g-solution}).

\vspace{9mm}

\setcounter{equation}{0}

\def\theequation{D.\arabic{equation}}

\noindent {\bf \large Appendix D}

\vspace{2mm}

\noindent In this appendix we present six pictures of spiky string
configurations at a fixed time. They are constructed by eq.
(\ref{X1-X_2=2}) for different values of $n$ and $\bar n,$ with
$\Lambda=|n\bar n|$. This value of the scale factor $\Lambda$ is
chosen just for convenience. The values on $n$ and $\bar n$ are
indicated below the pictures. The left pictures correspond to $n\bar
n<0$ and the rights to $n\bar n>0$. These pictures demonstrate the
properties 1-5, discussed in subsection 2.3.

\vspace{5mm}

\begin{figure}[h]
\begin{tabular}{cc}
\includegraphics[height=4.8cm]{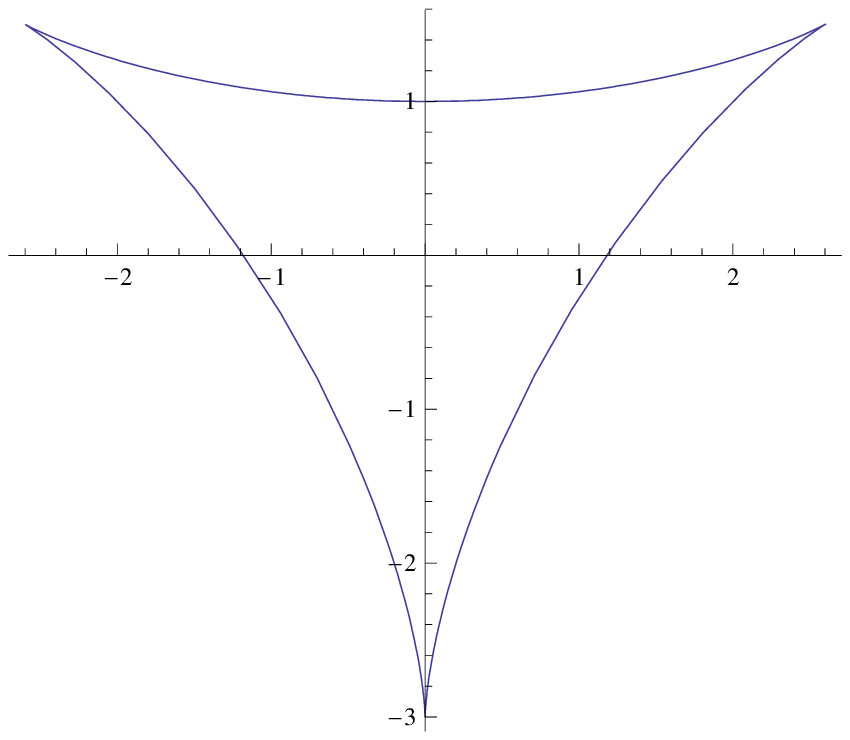}
\hspace{8mm}
\includegraphics[height=4.8cm]{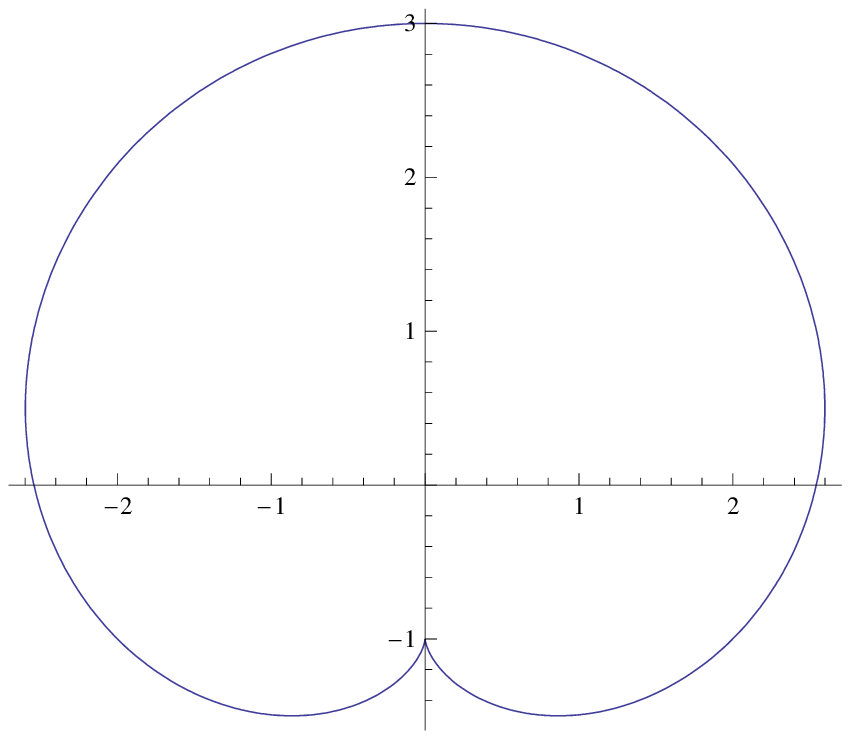}
\end{tabular}

\noindent ~~~~~~~~~~~~~~~{\bf Fig.$\,$1} {\it  $~~~n=2\,,~~$  $~\bar n=-1.$}
 $~~~~~~~~~~~~~~~~~~~~~~${\bf Fig.$\,$2} 
{\it  $~~~n=2\,,~~$  $~\bar n=1.$}
\end{figure}

\vspace{2mm}

\begin{figure}[h!!]
\begin{tabular}{cc}
\includegraphics[height=4.8cm]{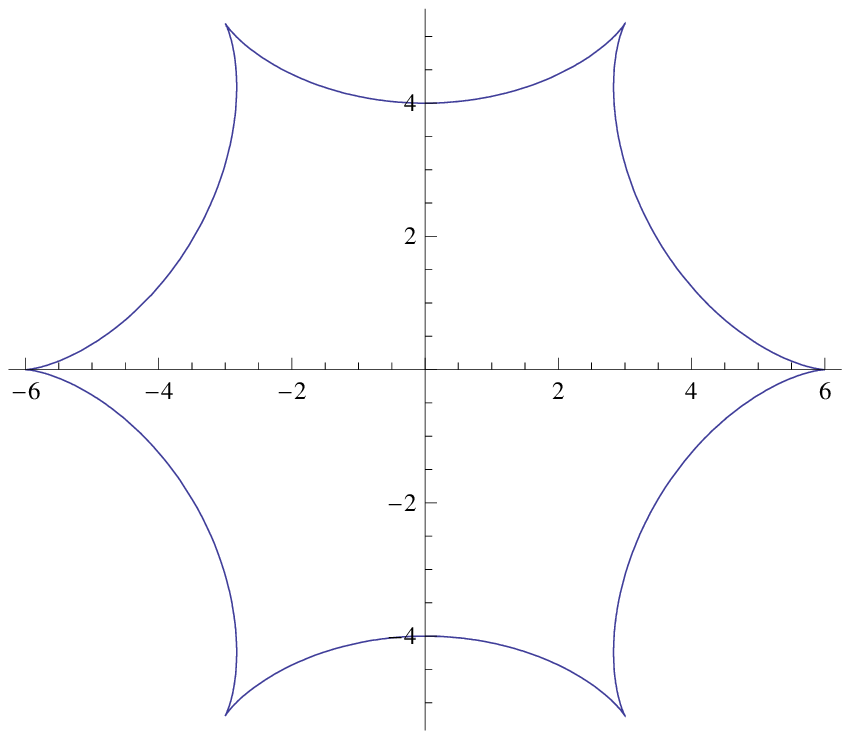}
\hspace{18mm}
\includegraphics[height=4.8cm]{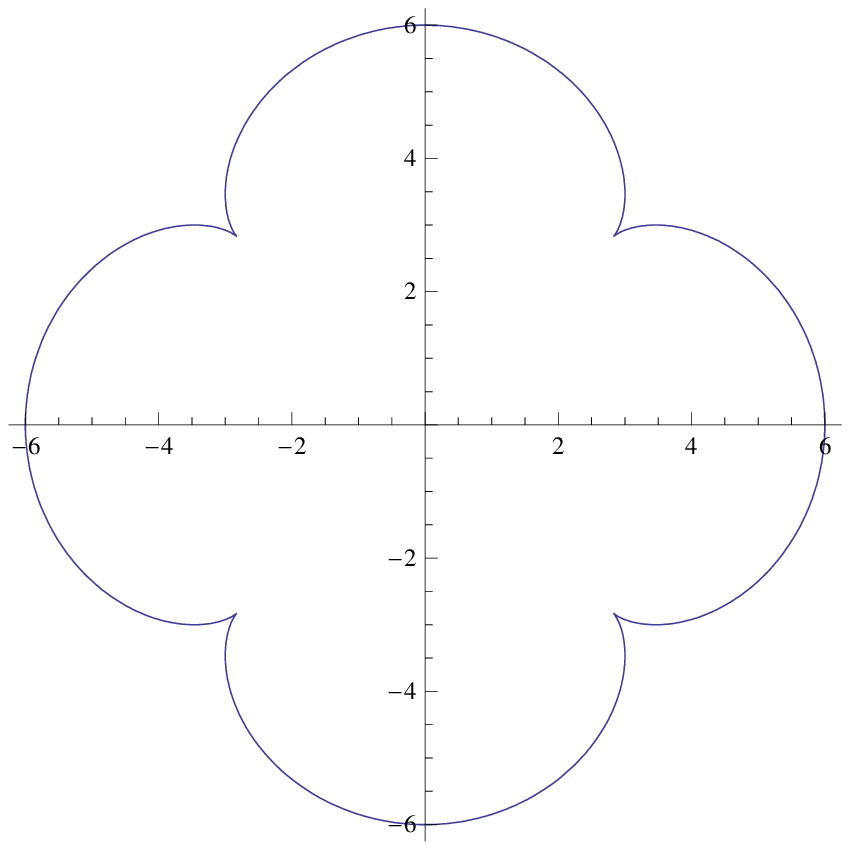}
~~~~~~~~~~
\end{tabular}

\noindent ~~~~~~~~~~~~~~
{\bf Fig.$\,$3} {\it  $~~~n=5\,,~~$  $~\bar n=-1.$}
~~~~~~~~~~~~~~~~~~~~~~~{\bf Fig.$\,$4} {\it  $~~n=5\,,~~$  $~\bar n=1.$}
\end{figure}

\vspace{2mm}

\begin{figure}[h!!]
\begin{tabular}{cc}
~~~~~~~~~~~~~~~~\includegraphics[height=4.8cm]{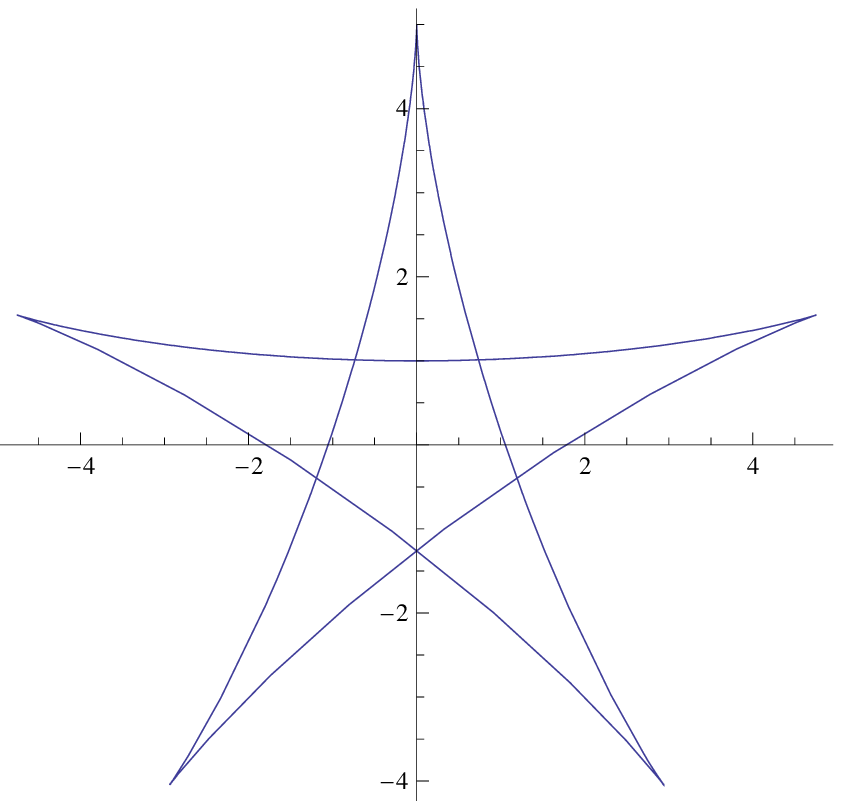}
\includegraphics[height=4.8cm]{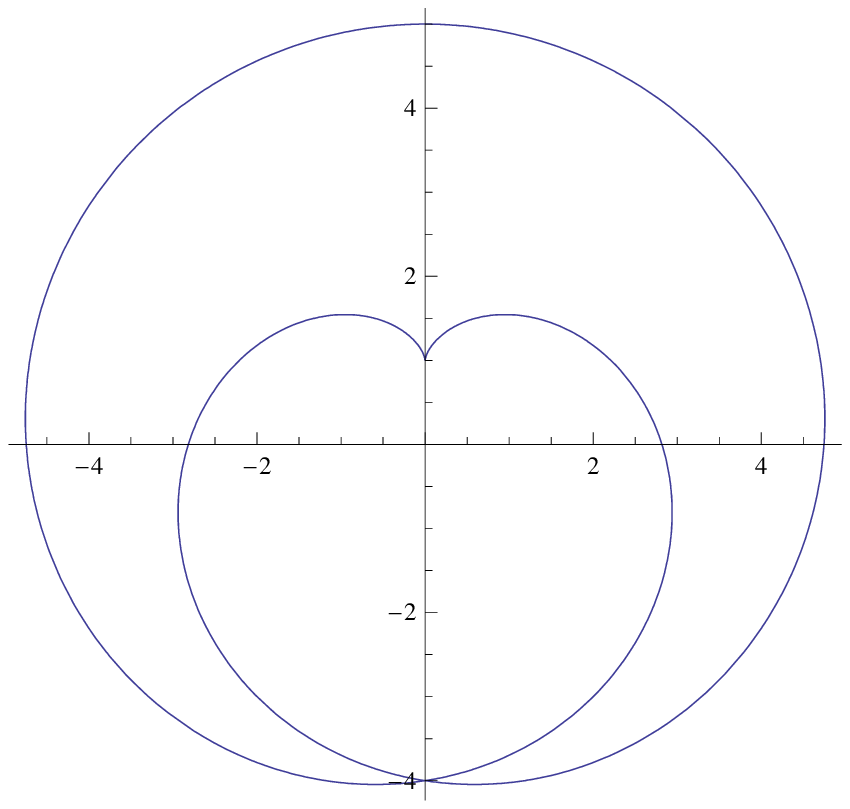}
\end{tabular}

\noindent $~~~~~~~$
~~~~~~~~~~~{\bf Fig.$\,$5} {\it  $~~~n=3\,,~~$  $~\bar n=-2.$}
 $~~~~~~~~~~~~~~~~~~~~${\bf Fig.$\,$6} {\it  $~~~n=3\,,~~$  $~\bar n=2.$}
\end{figure}

\vspace{9mm}

\newpage

\setcounter{equation}{0}

\def\theequation{E.\arabic{equation}}

\noindent {\bf \large Appendix E}

\vspace{3mm}

\noindent In this appendix we continue the discussion on spiky
string surfaces in $\rr^{1,2}$ and give a proof of the properties 3,
4 and 5 presented in subsection 2.3.

First note that the normal vector (\ref{N=F bar F}) can be written
in the form (\ref{N=B tims B}), with the conformal factor
(\ref{Liouv-g-solution}) and $B,$ $\bar B$ given by (\ref{B=F}).
Eqs. (\ref{B=()}) and (\ref{e^alpha=}) then yield
\begin{equation}\label{N=cos,sin}
N=\frac{1}{\cos\varphi_{_{-}}}\,
\left(\begin{array}{cr}~~\sin\varphi_{_{-}} \\ -\sin\varphi_{_{+}}\\
~~\cos\varphi_{_{+}}\end{array} \right)~,
\end{equation}
where $\varphi_{_{+}}$ and $\varphi_{_{-}}$ are the following
variables
\begin{equation}\label{varphi_pm}
\varphi_{_{\pm}}=\frac{1}{2}\,(nz\mp\bar n\bar z)=
\frac{1}{2}\,(n\mp\bar n)\tau+\frac{1}{2}\,(n\pm\bar n)\sigma~.
\end{equation}
The spikes (\ref{zeros of e^alpha}) obviously correspond to
$\cos\varphi_{_{-}}=0$ (or $|\sin\varphi_{_{-}}|=1$). Therefore, the
normal vector (\ref{N=cos,sin}) is singular at the spikes. Having
the unit norm, this vector diverges in the lightlike direction as in
the lightcone gauge.

The projection of the string surface (\ref{X=()}) on the
$(X_1,X_2)$-plane can be written as
\begin{equation}\label{X1-X_2=}
\vec X(\tau,\sigma)=\Lambda\, \frac{n+\bar n}{n\bar
n}\,\cos\varphi_{_{-}}
\left(\begin{array}{cr}~~\sin\varphi_{_{+}}\\
-\cos\varphi_{_{+}}\end{array}\right)-\Lambda\,\frac{n-\bar n}{n\bar
n}\,\sin\varphi_{_{-}}
\left(\begin{array}{cr}\cos\varphi_{_{+}}\\
\sin\varphi_{_{+}}\end{array}\right)~.
\end{equation}
Here  $\vec X=(X_1,X_2)$ denotes a 2d vector and $\varphi_{_{\pm}}$
are the variables (\ref{varphi_pm}). The module square of the
Euclidean vector (\ref{X1-X_2=}) is given by
\begin{equation}\label{X cdot X}
\vec X\cdot \vec X= \frac{\Lambda^2}{n^2\,\bar n^2}\left[n^2+\bar
n^2+2n\bar n\cos(2\varphi_{_{-}})\right]~.
\end{equation}
Since the spikes (\ref{zeros of e^alpha}) correspond to
$\cos(2\varphi_{_{-}})=-1$, they rotate around the origin on a
circle of the radius (\ref{spiky distance}). The differentiation of
(\ref{X1-X_2=}) yields
\begin{equation}\label{d_sigma X}
\partial_\tau{\vec X}=2\Lambda\,\sin\varphi_{_{-}}
\left(\begin{array}{cr} -\sin\varphi_{_{+}}\\
~~\cos\varphi_{_{+}}\end{array} \right)~, ~~~~~\partial_\sigma{\vec
X}=2\Lambda\,\cos\varphi_{_{-}}
\left(\begin{array}{cr}  \cos\varphi_{_{+}}\\
\sin\varphi_{_{+}}\end{array} \right)~.
\end{equation}
Expanding the tangent vector $\partial_\sigma\vec X(\tau,\sigma)$
for a fixed $\tau$ near to a spike and taking into account that
$\cos\varphi_{_{-}}=0$ at the spikes, from (\ref{X1-X_2=}) we find
\begin{equation}\label{X' approx}
\partial_\sigma\vec X(\tau,\sigma_m+\delta\sigma)
= n\bar n\, \vec
X(\tau,\sigma_m)\,\,\delta\sigma+O(\delta\sigma)^2~,
\end{equation}
where $\sigma_m$ is a solution of (\ref{zeros of e^alpha}) for a
fixed $\tau\,$ and $\,\delta\sigma$ is an infinitesimal variation.
Thus, the vector $\partial_\sigma X(\tau,\sigma)$ inverts its
direction at $\sigma=\sigma_m$, which indicates on the spiky
character of the singularity. Due to (\ref{X' approx}), the spike at
$\sigma_m$ and the radius vector $\vec X(\tau,\sigma_m)$ have the
same or opposite directions, depending on the sign of $n\bar n,$ as
it is stated in the item 3.

Let's consider the curve $\vec X(\tau,\sigma)$ as a function of
$\sigma$ for a given $\tau$. Its curvature is defined by the normal
(to $\partial_\sigma\vec X$) component of
$\partial^2_{\sigma\sigma}\vec X$
\begin{equation}\label{X"_N}
\partial^2_{\sigma\sigma}\vec X|_{_{N}}=
\partial^2_{\sigma\sigma}\vec X-
\frac{\partial^2_{\sigma\sigma}\vec X\cdot\partial_\sigma\vec
X}{\partial_\sigma\vec X\cdot\partial_\sigma\vec X}\,\,
\partial_\sigma\vec X~.
\end{equation}
The curvature with respect to the origin is positive if the scalar
product $\partial^2_{\sigma\sigma}\vec X|_{_{N}}\cdot\vec X$ is
negative and vice versa. Calculating the vector
$\partial^2_{\sigma\sigma}\vec X$ from (\ref{d_sigma X}) and using
(\ref{X"_N}), we obtain
\begin{equation}\label{X"_N cdot X}
\partial^2_{\sigma\sigma}\vec X|_{_{N}}\cdot\vec X=-\Lambda^2\,
\frac{(n-\bar n)^2}{n\,\bar n}\,\,\cos^2\varphi_{_{-}}~.
\end{equation}
This equation proves the property 4 for an arbitrary $\tau$.

Finally, considering the property 5, we calculate the differential
of the polar angle for the curve (\ref{X1-X_2=2}) and integrate it
in the interval $[\sigma_m,\sigma_{m+1}]$ defined by
(\ref{sigma_m}). This provides the rotation angle
\begin{equation}\label{delta phi=}
\Delta\phi=\frac{|n\bar n(n+\bar n)|}{|n-\bar
n|}\,\int_0^{2\pi}d\theta\,\,\frac{1+\cos\theta}{n^2+\bar n^2+2n\bar
n\cos\theta}~,
\end{equation}
where $\theta=|n-\bar n|\sigma$. Writing (\ref{delta phi=}) as a
contour integral over the unit circle with $\zeta=e^{i\theta},\,$
and calculating the residues of the integrand at the poles inside
the unit disk we obtain (\ref{delta phi}). Note that the integrand
has three poles at $\zeta_1=0,$  $\zeta_2=-n/\bar n,$
$\zeta_3=-\bar n/n$ and only two of them are inside the unit disk.

\vspace{9mm}

\setcounter{equation}{0}

\def\theequation{F.\arabic{equation}}

\noindent {\bf \large Appendix F}

\vspace{3mm}

\noindent Here we present some useful formulas for $sl(2,\rr)$
algebra and $SL(2,\rr)$ group.

From eq. (\ref{tt=}) follows $a^2=\langle a\,a\rangle\,I,$ for any
$a\in sl(2,\rr)$. In particular, if $\langle a\,a\rangle=0$, then
$a$ is nilpotent $a^2=0$. Eq. $a^2=\langle a\,a\rangle\, I\,\,$
helps to find a compact form
of $\,e^a$
\begin{eqnarray}\label{e^A}
e^{a}=\cosh\theta\,\,I+{\sinh\theta}\,\,\hat a~,~~\mbox{with}~~~
\theta=\sqrt{\,\langle\,\,a\,a\,\rangle}~,~~~~\hat
a=\frac{a}{\theta}~, ~~~~\mbox{if}~~~\langle\,a\,a\,\rangle
>0~;~~~
\\ \nonumber
e^{a}=\cos\theta\,\,\,I\,+\,{\sin\theta}\,\,\,\hat
a~,~~~~\mbox{with}~~~ \theta=\sqrt{-\langle\,a\,a\,\rangle}~,~~~\hat
a=\frac{a}{\theta}~,
~~~~\mbox{if}~~~ \langle\,a\,a\, \rangle <0~;~~~\\ \nonumber
e^{a}=I +a~,~~~~~~~~~~~~~
~~~~~~~~~~~~~~~~~~~~~~~~~~~~~~~~~~~~~~~~~~~~~~~~~~~~~\mbox{if}~~~
\langle\, a\,a\, \rangle =0~.~~~
\end{eqnarray}
From these equations follows that
\begin{eqnarray}\label{<e^A>}
\langle\,e^{a}\rangle >1~, ~~~~\mbox{if}~~~\langle\,a\,a\,\rangle >0~;~~~~
(\mbox{spacelike}~~a~~\mbox{and hyperbolic}~~e^{a}) \\ \nonumber
\langle\,e^{a}\rangle \in (-1,1)~,
~~~~\mbox{if}~~~ \langle\,a\,a\, \rangle <0~;~~~~
(\mbox{timelike}~~a~~\mbox{and elliptic}~~e^{a})\\ \nonumber
\langle\,e^{a}\,\rangle =1~,~~~~\mbox{if}~~~
\langle\, a\,a\, \rangle =0;.~~~~
(\mbox{lightlike}~~a~~\mbox{and parabolic}~~e^{a})
\end{eqnarray}
The $SO(2)$ subgroup of $SL(2,\rr)$ given by
\begin{equation}\label{e^T}
e^{\theta\, t_0}=\left( \begin{array}{cr}~ \cos\theta&\sin\theta\\
-\sin\theta&\cos\theta
\end{array}\right)~
\end{equation}
defines the rotations of $t_1$ and $t_2$ under the transformations
of the adjoint representation
\begin{equation}\label{e^-t e^t}
e^{\frac{1}{2}\,\theta\, t_0}\,t_1\,e^{-\frac{1}{2}\,\theta\,
t_0}=t_1\,\cos\theta+t_2\,\sin\theta~,~~~~~~
e^{\frac{1}{2}\,\theta\, t_0}\,t_2\,e^{-\frac{1}{2}\,\theta\,
t_0}=t_2\,\cos\theta-t_1\,\sin\theta~.
\end{equation}
Other useful relations are
\begin{equation}\label{(t1+it2)e^t}
(t_1+it_2)\,e^{\theta\,t_0}=e^{i\theta}\,(t_1+it_2)=e^{-\theta\,t_0}\,(t_1+it_2)~,
\end{equation}
\begin{equation}\label{(I+it0)e^t}
(I-it_0)\,e^{\theta\,t_0}\,=\,e^{i\theta}\,(I-it_0)\,=\,e^{\theta\,t_0}
\,\,(I-it_0)~.
\end{equation}


\end{document}